\documentclass[english,journal]{IEEEtran}
\usepackage[T1]{fontenc}
\usepackage[latin9]{inputenc}
\usepackage{color}
\usepackage{array}
\usepackage{units}
\usepackage{multirow}
\usepackage{amsmath}
\usepackage{amsthm}
\usepackage{amssymb}
\usepackage{graphicx}

\makeatletter

\providecommand{\tabularnewline}{\\}

\theoremstyle{definition}
\newtheorem{defn}{\protect\definitionname}
\theoremstyle{plain}
\newtheorem{thm}{\protect\theoremname}
\theoremstyle{plain}
\newtheorem{cor}{\protect\corollaryname}
\theoremstyle{plain}
\newtheorem{prop}{\protect\propositionname}

\@ifundefined{date}{}{\date{}}

\usepackage{amsfonts}
\usepackage{cite}
\usepackage{array}
\usepackage{algorithm}
\usepackage{algorithmic}
\usepackage{subfigure}
\usepackage{stfloats}
\usepackage{graphicx}

\usepackage[font=small,labelfont=bf]{caption}

\usepackage{gensymb}

\makeatother

\usepackage{babel}

\makeatother

\usepackage{babel}
\providecommand{\corollaryname}{Corollary}
\providecommand{\definitionname}{Definition}
\providecommand{\propositionname}{Proposition}
\providecommand{\theoremname}{Theorem}

\begin{document}
\title{Analysis of Channel Uncertainty in Trusted Wireless Services via Repeated
Interactions}
\author{Bingwen Chen, Xintong~Ling,~\IEEEmembership{Member,~IEEE}, Weihang~Cao,~\IEEEmembership{Graduate
Student Member,~IEEE}, \\
Jiaheng~Wang,~\IEEEmembership{Senior Member,~IEEE}, Zhi~Ding,~\IEEEmembership{Fellow,~IEEE}
}
\maketitle
\begin{abstract}
The coexistence of heterogeneous sub-networks in 6G poses new security
and trust concerns and thus calls for a perimeterless-security model.
Blockchain radio access network (B-RAN) provides a trust-building
approach via repeated interactions rather than relying on pre-established
trust or central authentication. Such a trust-building process naturally
supports dynamic trusted services across various service providers
(SP) without the need for perimeter-based authentications; however,
it remains vulnerable to environmental and system unreliability such
as wireless channel uncertainty. In this study, we investigate channel
unreliability in the trust-building framework based on repeated interactions
for secure wireless services. We derive specific requirements for
achieving cooperation between SPs and clients via a repeated game
model and illustrate the implications of channel unreliability on
sustaining trusted wireless services. We consider the framework design
and optimization to guarantee SP-client cooperation, given the worst
channel condition and/or the least cooperation willingness. Furthermore,
we explore the maximum cooperation area to enhance service resilience
and reveal the trade-off relationship between transmission efficiency,
security integrity, and cooperative margin. Finally, we present simulations
to demonstrate the system performance over fading channels and verify
our results.

\vspace{-0.5cm}XINLAB2025
\end{abstract}

\section{Introduction}

6G is envisioned to be an open and integrated network that enables
the coexistence of heterogeneous sub-networks and diverse terminals
for providing seamless coverage across space, air, ground, and sea
\cite{You2021}. Clients should not be restrained to the subscribed
service provider (SP) but be able to choose the most proper one from
diverse SPs. However, on the one hand, the booming and complicated
heterogeneous sub-networks make the verification of SPs' identity
difficult; on the other hand, the existence of low reliable devices,
such as drones and smart sensors, makes the system vulnerable and
weakens the trust among different entities. Traditional security models
typically use a \textquotedbl trust but verify\textquotedbl{} approach
\cite{Dhiman2024}, relying on static, centralized techniques such
as firewall \cite{Nahar2024}, intrusion detection system (IDS) \cite{Hajj2021},
and single sign-on (SSO)\cite{Mahnamfar2024}. Even though some federated
trust models enable authentication sharing across different organizations
\cite{Gao2010}, they still adopt the same principle as SSO that requires
explicit network boundaries or perimeters and assumes users to be
trustworthy once authenticated by authorities\cite{Kang2023}.

Due to the limitations of the above perimeter-based security models,
large-scale open 6G networks call for a novel perimeterless zero-trust
architecture, which is based on the main motto of \textquotedbl never
trust, always verify\textquotedbl{} \cite{Buck2021} and has been
introduced into the areas such as Internet of Things \cite{Samaniego2018},
edge computing \cite{Sedjelmaci2024}, mobile networks \cite{Liu2024},
to name just a few. In zero-trust architecture, all entities, including
internal ones, are not trusted by default, and every access request
requires dynamic identity and access validation \cite{Baseri2018,Dimitrakos2020,Hatakeyama2021}.
A typical example is Google's BeyondCorp \cite{Ward2014}, which continuously
verifies user and device trustworthiness when accessing, based on
device state and user credentials. However, the development of zero-trust
architecture is still in its infancy. The continuous identity verification
request not only increases the authentication overhead but also relies
on a centralized and static model, which limits the scalability and
adaptability of next-generation network (NGN). 

Blockchain technology offers a novel approach toward zero trust by
leveraging its transparency, anonymity, and decentralization \cite{Zheng2018}.
For example, blockchain radio access network (B-RAN), proposed in
\cite{Ling2019,Le2019,Ling2020a}, is a distributed wireless access
paradigm enabling multi-sided cooperation across sub-networks without
pre-established trust or centralized authority. As a multi-sided platform,
B-RAN integrates distributed resources owned by different sub-networks
and allows clients to access services beyond the specific subscribed
SPs \cite{Ling2021}. However, the trust establishment between clients
and SPs still faces a significant challenge to supervise the quality
and experience of off-chain services\cite{Ling2020a,Ling2021b}, severely
weakening the trust foundation of B-RAN, especially in a zero-trust
environment. 

In the context of B-RAN, the authors in \cite{cao2024} designed a
wireless access framework by segmenting the entire service into multiple
rounds and modeling the repeated interactions between SPs and clients
by a sequential game. The SP first provides a trial service, during
which the client must make a payment. Upon receiving the payment,
the SP continues to provide the remainder of the service for that
round. This process is repeated until the whole service is completed.
Through the interactions, both parties continuously verify each other's
trustworthiness and maintain the service until detecting any dishonest
actions. The repeated interactions can be viewed as the trust-building
process. With proper system configuration, participants are motivated
by long-term profits to regulate their behavior and adopt honest strategies
to ensure the off-chain service quality and thus address the above
challenges in B-RAN. Similar methods have been investigated in \cite{Antoniou2011,Kamhoua2012,Wang2019}
to establish cooperation among different network entities. Among these
works, participants keep monitoring the actions of their opponents
and continuously evaluate the trustworthiness between them, which
aligns with the principle of zero trust, i.e., continuous monitoring
and validation. 

However, the existing works have overlooked the impact of unpredictable
environmental uncertainty on the trust-building process relying on
continuous, repeated interactions. During wireless services, wireless
communication channels between the SPs and the clients are unreliable
and may experience outages due to channel fading. Usually, the channel
outage cannot be distinguished from a deliberately defection by any
party. In the repeated game, such situations are more severe and may
even result in perceptual errors \cite{Axelrod1988}, where both parties
mistakenly believe the other has defected. As a result, even if both
parties are honest, the service process may still terminate because
of the physical channel outage. Due to the inevitability of channel
fading in practice, such a trust-building process could be vulnerable
and unstable. 

\textcolor{red}{}Therefore, in this work, we consider a critical
problem of how environmental uncertainty affects the trust-building
process based on repeated interactions.\textcolor{red}{{} }Specifically,
we would like to seek the potentiality of achieving cooperation via
repeated interactions over unreliable channels and characterize the
cooperation conditions by taking channel uncertainty into consideration.
Based on the cooperation conditions, we would like to optimize the
trust-building process under channel uncertainty and reveal the potential
trade-off relationships therein. Remark that our work aligns well
with the trust notion in \cite{Ling2025} where trustworthiness is
characterized by two key dimensions: commitment and competence. Specifically,
in our context, the competence is the channel reliability, whereas
the commitment is the willingness to cooperate. Our work can be viewed
as a typical example to illustrate how the competence and commitment
impact the trust-building process. 

The key contributions of this work are listed as follows:
\begin{itemize}
\item We establish a repeated game model to describe the interactive trusted
wireless services over unreliable channels and illustrate the impact
of channel uncertainty on the long-term payoffs of different strategies. 
\item We define the cooperation conditions under which the honest cooperation
strategy is dominant among the other strategies including dishonest
actions. We derive the specific requirement for achieving cooperation
and assess the impact of channel unreliability on maintaining interactive
trusted access services.
\item Given a worst-case channel condition, we formulate the optimization
problem to guarantee the SP-client cooperation with the least requirement
on cooperation willingness. We also show the minimum requirement on
channel quality, given the worst-case cooperation willingness of both
parties. Interestingly, our result reveals that the optimal solutions
to the above two problems can be achieved simultaneously.
\item We introduce the concept of cooperation region, defined as the set
of cooperation willingness and channel outage probability that can
guarantee cooperation between the SP and the client. A larger cooperation
region implies a stronger robustness of our framework. By considering
the cooperation area maximization, we point out that our framework
should take both parties into consideration and would be problematic
if the trust-building process highly relies on one side.\textcolor{magenta}{{} }
\item Furthermore, we analyze the trusted framework from three aspects:
transmission efficiency, service integrity, and cooperation margin,
and uncover the trade-off relationship among them, which provides
more comprehensive guidance and insights for practical design.
\item We show the performance of the interactive trusted wireless services
over fading channels via simulations. Our results highlight how channel
unreliability and cooperation willingness affect the service process
based on repeated interactions and offer insights on the corresponding
trust-building process.\textcolor{magenta}{}
\end{itemize}
The rest of the paper is organized as follows. Section \ref{sec:access}
presents the system model. Section \ref{sec:gamemodel} establishes
a repeated game for different strategies. Section \ref{sec:coopCond}
derives the cooperation conditions. Section \ref{sec:SystemConfOpt}
considers the framework design under cooperation conditions. Section
\ref{sec:CoopArea} maximizes the cooperation region, and Section
\ref{sec:TradeoffAnalysis} analyzes the trade-off relationships therein.
Section \ref{sec:Simulations} provides simulation results and Section
\ref{sec:Conclusion} concludes the paper.\vspace{-0.3cm}

\section{Trusted Wireless Services via Repeated Interactions\label{sec:access}}

\subsection{Framework Overview\label{subsec:Protocol}}

\begin{figure}
\begin{raggedright} \centering\includegraphics[viewport=0bp 0bp 566bp 504bp,width=0.5\textwidth]{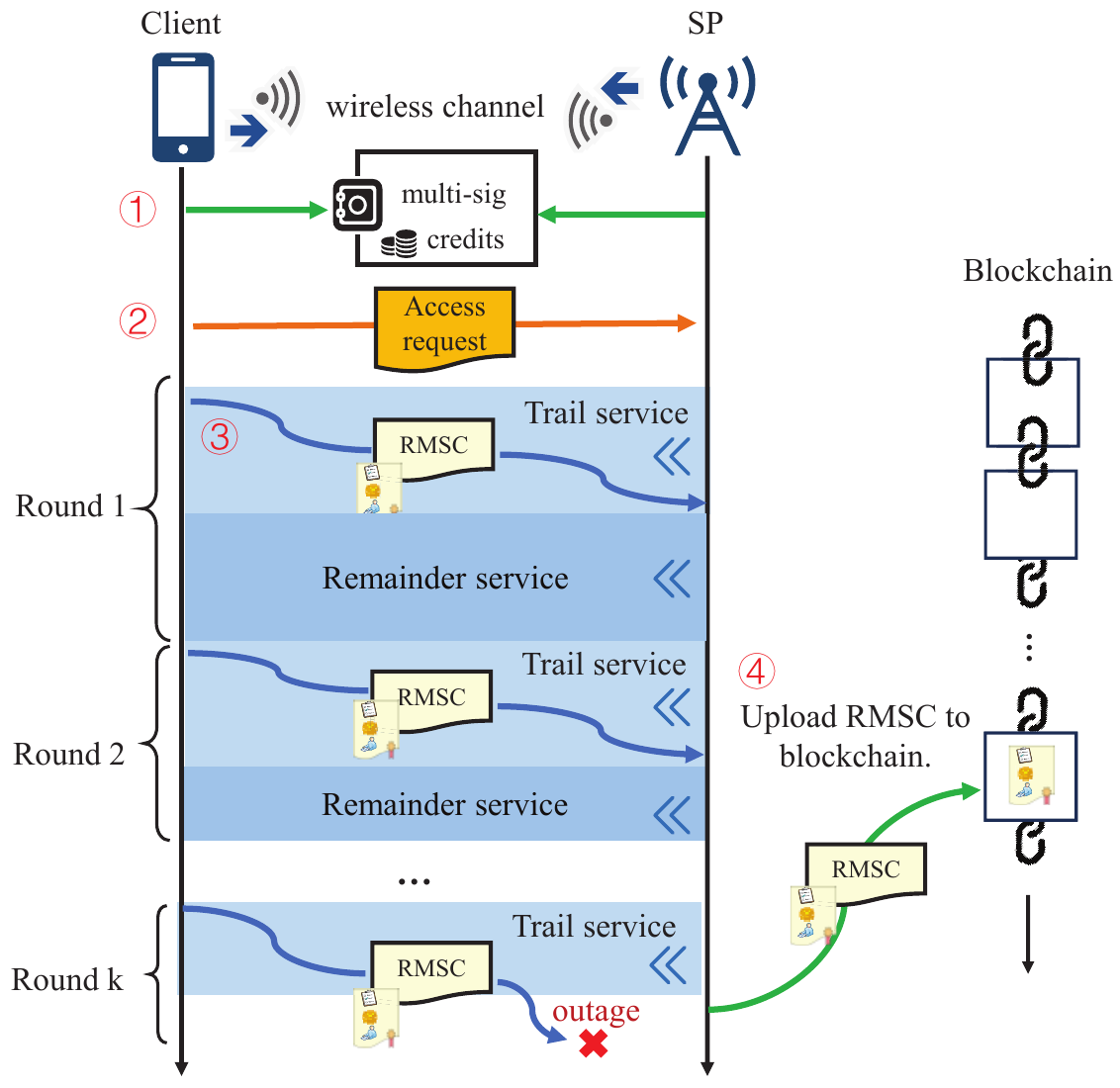}

\end{raggedright} \centering{}

\caption{Workflow of trusted wireless access services via repeated interactions.\label{fig:workflow}}
\vspace{-0.6cm}
\end{figure}

This section presents the trusted wireless access framework based
on repeated interactions in detail. Consider a zero-trust wireless
environment without pre-established trust. The absence of authentication
may lead the network participants to behave selfishly and deceive
each other for their own benefit. For instance, payment-before-service
is widely adopted in centralized schemes, in which the SP and the
client first negotiate on the price, time, and quality of service
(QoS), and then the service is provided after the client pays the
bill. However, in such an untrustworthy environment, after receiving
payment, the SP may fail to provide the negotiated service due to
objective reasons (e.g., terrible channel conditions or interference)
or subjective ones (e.g., the SP deliberately selects a bad channel
or reduces the transmission power). For either reason, the trust between
the client and the SP is significantly weakened.

To tackle the above issue, the trust-building framework in \cite{cao2024}
establishes trusted wireless access services between an SP and a client
via repeated interactions. The main idea is that trust is built through
successive interactions and continuous evaluation of each other's
behavior. Fig. \ref{fig:workflow} illustrates the detailed procedure
as follows. 
\begin{itemize}
\item \textbf{Step1.} Before the service delivery, a lightning channel between
the client and the SP is established by funding a multi-signature
address with sufficient credits through an enhanced hashed timelock
contract (eHTLC) \cite{Le2021}. 
\item \textbf{Step2.} The service starts after the SP and the client negotiate
the service terms and write them down in the access request. We denote
the service time by a random variable $T$. The whole service is segmented
into multiple rounds with the same duration, i.e., the slot time denoted
by $t$. Given a service with duration $T$, the whole service requires
$M=\lceil T/t\rceil$ rounds to finish.
\item \textbf{Step3.} Each round starts with a period of trial service provided
by SP. During the trial time which is denoted by $\tau$, clients
need to pay for the service fee of the current round to the SP by
using a special smart contract named revocable sequence maturity contract
(RSMC). Once the trial service is completed and the client's payment
is received, SP continues with the remainder services; otherwise,
service provision is halted. This cycle repeats each round until the
entire service is fulfilled or interrupted by dishonest behaviors.
In the process, we denote the SP\textquoteright s cost for providing
service per unit of time as $c$ and the price charged per unit of
time as $p$. Note that the client's payment via smart contract may
fail due to the channel outage, which will be discussed in Section
\ref{subsec:Communication-Model}.
\item \textbf{Step4. }After the service is terminated, the SP uploads all
the RSMCs on-chain, which will update the eHTLC balance of the client
and the SP. The payments are then verified by the miners in the blockchain
network, which is asynchronous with the access process so that wireless
access services do not need to wait for verification.
\end{itemize}
The trusted wireless access via interactions aligns with the zero-trust
principle of \textquotedbl never trust, always verify\textquotedbl .
On the side of ``never trust'', it does not require any pre-established
trust from trusted third parties, and thus is suitable for distributed
wireless networks. On the other side of \textquotedbl always verify\textquotedbl ,
the trust-building process is based on continuous evaluation of the
other's actions. For the client, once detecting the dishonest behavior
of the SP, e.g., the service fails to meet the expected QoS, the client
would not pay for the current round and may even terminate the service,
incentivizing the SP to maintain high-quality service for long-term
benefits. Conversely, if the client does not pay during the trial
time, the SP will cease to provide further service, which also encourages
the client to cooperate to secure long-term profits. To maximize the
long-term benefits, both parties are incentivized to act honestly
for long-term cooperation, which fosters an interest-driven trust
between the client and the SP. 

We would like to highlight that the above framework is scalable for
distributed 6G networks. First, it does not rely on any centralized
authorities, which is often a critical scalability bottleneck in traditional
zero-trust architectures \cite{Alevizos2022}. The distributed trust-building
process reduces the complexity of trust management as the network
scales. Second, we use eHTLC to build off-chain payment channels to
reduce the scalability bottleneck caused by blockchain in managing
frequent micro-transactions during repeated interactions. Similar
to most layer-2 technologies \cite{Papadis2020}, eHTLC does not require
real-time consensus across the entire network to avoid very lengthy
block confirmation latency.\vspace{-0.3cm}

\subsection{Communication Model\label{subsec:Communication-Model}}

In \cite{cao2024}, clients and SPs are assumed to communicate through
an ideal channel with no fading, i.e., QoS is determined by the SP's
willingness. However, wireless channels are subject to fading over
time and may experience outages. Deep fading can result in the failure
to recover packets at the receiver and lower the QoS. Such channel
uncertainty is independent of the SP's willingness, and always exists
in practice, even though both the SP and the client are willing to
cooperate.\textcolor{red}{} \textcolor{red}{}

We denote the channel outage probability by $d$, which specifically
represents the probability of a packet being lost due to channel fading.
Note that channel fading is the inherent characteristic of wireless
channels that cannot be completely avoided. However, it is possible
to influence $d$ via adjusting, e.g., coding rate, modulation scheme,
transmission power, or transmission frequency. In our work, the outage
probability $d$ is assumed to be quasi-constant during a single service
process. 

Channel unreliability significantly affects the interests of both
clients and SPs. For clients, the outage probability $d$ reflects
the channel quality and influences the client's service experience.
We use a utility function $\Gamma\left(d\right)$ to characterize
the client's interest in receiving service usage per unit time. Without
loss of generality, the utility function $\Gamma\left(d\right)$ is
monotonically decreasing with $d$, indicating that packet loss degrades
quality of experience (QoE) and utility, with $\Gamma\left(0\right)$
representing the maximum utility over an ideal channel with no outage.
We also assume $\Gamma\left(d\right)$ to be economically feasible,
i.e., $\Gamma\left(0\right)>p$; otherwise, the service is not financially
worthwhile for the client. Most of our analysis is merely based on
the two assumptions of the utility function $\Gamma\left(d\right)$.
We would like to present several typical examples of the utility functions
such as the linear utility function $\Gamma\left(d\right)=\Gamma\left(0\right)\left(1-d\right)$.
If we consider diminishing marginal utility\cite{Besanko2020}, the
utility function can be formulated by $\Gamma\left(d\right)=\Gamma\left(0\right)\exp\left(-\varphi d\right)$,
where $\varphi>0$ represents the client's sensitivity to outage.
A larger $\varphi$ indicates a more sensitive client to the service
quality. One can see the high channel outage harms the client's interest
and potentially influences the client's willingness to pursue further
services from the SP. 

Channel outage impacts not only clients but also SPs. Recall that,
in our framework, the SP expects to receive payment each round through
transactions based on smart contracts. Hence, the client must transmit
a packet containing transactions each round, but these packets are
also subject to channel outage with the probability $d$. Consequently,
the SP may fail to recover the transaction and thus cannot get the
corresponding credits. In this case, the SP cannot determine whether
the packet loss is due to channel fading or the client's dishonest
behavior. The service process would terminate due to the channel fading,
even when both parties are honest and willing to cooperate. Hence,
the channel quality subtly impacts both sides. If the SP deliberately
lowers QoS, it may also hurt its own interest. 

In conclusion, although a relatively simple channel model is considered,
it properly characterizes physical environmental uncertainty within
the considered framework. This channel unreliability, independent
of the willingness of SPs or clients, is objective and significantly
impacts the service process and cooperation between SPs and clients,
which are subjective and depend on their willingness. Imagine that
the SP and the client are willing to cooperate, but the payment in
the smart contract fails to be captured by the SP due to channel fading.
In this case, SP may assume the client is at fault and cease to provide
services. Conversely, the client may believe the SP received the transaction
but failed to deliver the remaining services. Hence, both parties
may mistakenly presume the other is at fault, leading to the termination
of cooperation, even if they are willing to continue the service.
In this case, the trust perceived by SP or the client is no longer
the same as the actual one between them, which is affected by channel
outage. Therefore, it calls for an analytical characterization of
how channel uncertainty influences the zero-trust architecture based
on repeated interactions.

\vspace{-0.3cm}

\section{Repeated Game Modeling\label{sec:gamemodel}}

\subsection{One-round Payoff \label{subsec:PayoffAnalysis}}

In this section, we model the framework via game theory. Each round
in the service process is a stage game in a repeated game and is also
a sequential game, i.e., the SP and the client make a move sequentially,
being aware of the opponent's previous move. Denote the action profile
in a stage game by $\left(\sigma_{c},\sigma_{s}\right)$, with $\sigma_{c}$
and $\sigma_{s}$ representing the action played by the client and
the SP in each stage game, respectively. The payoffs of the SP and
the client in the stage game are denoted by $u_{s}\left(\sigma_{c},\sigma_{s}\right)$
and $u_{c}\left(\sigma_{c},\sigma_{s}\right)$, respectively. 

\begin{figure}
\begin{raggedright} \centering\includegraphics[viewport=0bp 0bp 346bp 440bp,clip,width=0.4\textwidth]{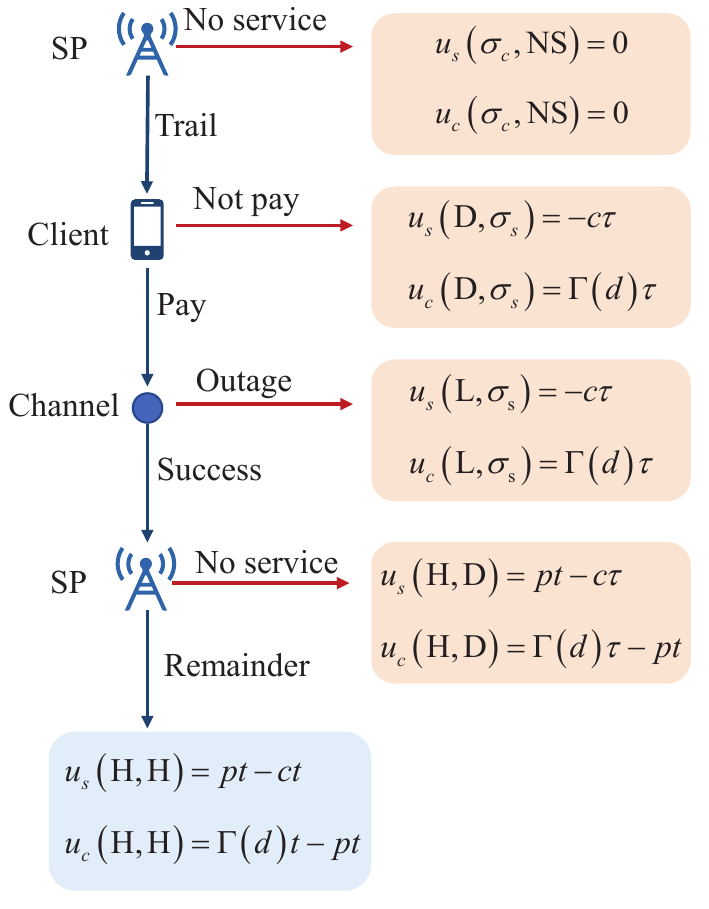}

\end{raggedright} \centering{}

\caption{Game tree for a stage game in repeated interactions.\label{fig:game tree}}
\vspace{-0.6cm}
\end{figure}
At the beginning of the round, if the SP decides not to provide any
service, namely \emph{No Service} ($\text{NS}$), both parties receive
no payoff with no further interactions, i.e., $u_{s}\left(\sigma_{c},\text{NS}\right)=u_{c}\left(\sigma_{c},\text{NS}\right)=0$.
Alternatively, if the SP provides the trial service, the client then
decides between paying before the trial is due (\emph{Honest} ($\text{H}$)),
or not paying (\emph{Defection} ($\text{D}$)). If the client defects,
the payoffs would be $u_{s}\left(\text{D},\sigma_{s}\right)=-c\tau$
and $u_{c}\left(\text{D},\sigma_{s}\right)=\Gamma\left(d\right)\tau$.
However, even if the client chooses \emph{Honest}, the credit exchange
may fail due to the channel outage, i.e., \emph{Loss} ($\text{L}$).
This results in the same payoffs as defection, i.e., $u_{s}\left(\text{\text{L}},\sigma_{s}\right)=-c\tau$
and $u_{c}\left(\text{\text{L}},\sigma_{s}\right)=\Gamma\left(d\right)\tau$,
indicating that the packet loss can financially benefit the client
while penalizing the SP in the short term. If the client is honest
and the credit exchange transaction successfully reaches the SP, the
SP must then decide whether to provide the remainder service or not,
namely \emph{Honest} ($\text{H}$) or \emph{Defection} ($\text{D}$).
If the remainder service is provided, the payoff pair is given by
$u_{s}\left(\text{\text{H}},\text{H}\right)=pt-ct$ and $u_{c}\left(\text{H},\text{H}\right)=\Gamma\left(d\right)t-pt$;
otherwise, it would be $u_{s}\left(\text{\text{H}},\text{D}\right)=pt-c\tau$
and $u_{c}\left(\text{H},\text{D}\right)=\Gamma\left(d\right)\tau-pt$.
The whole process can be illustrated by a game tree, shown in Fig.
\ref{fig:game tree}. 

In this model, participants are also allowed to compensate for the
defection and restore the previous mutual trust. For example, the
SP provides the remainder services that were undelivered in the previous
defection round, or the client repays for the unpaid service. Therefore,
besides the basic actions, the participants can adopt \emph{Recovery}
($\text{R}$) in a round following a defection. The payoffs of the
recovery round with the opponent being honest can be given by: $u_{s}\left(\text{R},\text{H}\right)=-c\left(t-\tau\right)$
and $u_{c}\left(\text{H},\text{R}\right)=pt-\Gamma\left(d\right)\left(t-\tau\right)$. 

\vspace{-0.3cm}

\subsection{Long-term Payoff\label{subsec:LongtermPayoff}}

For a one-shot game with only one round, one can easily observe through
backward induction that the SP would not provide any service since
the very beginning, and trusted access cannot be established. If the
interactions between the SP and the client are played multiple rounds,
then the framework is a non-zero-sum repeated game. In a repeated
game, players may choose different equilibrium strategies at the expense
of immediate interests for long-term interests \cite{Wu1995}. 

This subsection derives the long-term payoff for both the SP and the
client. We first introduce the probability of continuation, denoted
as $w$, which implies the probability of the $k+1$ round exists
after the $k$ round, and $w$ can be given by
\begin{align}
w & =\text{Pr}\left\{ M\geq k+1|M\geq k\right\} .\label{eq:wdef}
\end{align}
Then the long-term payoff of a repeat game is given by
\begin{equation}
\pi=\sum_{i=0}^{\infty}w^{i}u=\frac{u}{1-w},\label{eq:longTermPayoff0}
\end{equation}
where $\pi$ and $u$ represent the long-term and single-round payoff,
respectively.

On one hand, $w$ acts as the \emph{discount factor} in the repeated
game, which quantifies the conversion of future payoffs into present
value \cite{Fudenberg1991}. A player may benefit from a single round
of betrayal but at the expense of potential long-term returns. On
the other hand, $w$ can be considered as the \emph{continuance intention},
in that a higher $w$ implies that the participants are more willing
to engage in the service. Furthermore, $w$ also represents\textbf{
}\emph{cooperation willingness}, suggesting that participants believe
sustained cooperation leads to greater overall benefits, which encourages
them to apply cooperative strategies.

The above trust-building process is related to two factors: the environmental
uncertainty, which is objective and determined by the channel outage
$d$, and the cooperation willingness $w$, which is subjective. Remark
that they align with the kernel concept of trust: the competence,
i.e., the ability to achieve the cooperation in our context, and the
commitment, i.e., the intention to do so \cite{Ling2025}. Therefore,
the analysis in our work may provide meaningful insights into the
more generic notion of trust. Next, we will illustrate the specific
impact of these two factors. 

\vspace{-0.3cm}

\subsection{Strategy Analysis\label{subsec:strategyAnalysis}}

A repeated game requires more sophisticated cooperative strategies.
Now we introduce the cooperation strategy named \emph{COOP}, where
players behave honestly at the beginning until they detect the opponent\textquoteright s
betrayal behavior, and then stop cooperation until the other restores
honesty in a later round. When both players adopt COOP, however, a
perceptual error \cite{Axelrod1988} may occur if the credit transaction
is lost due to the channel fading, where both parties mistakenly believe
that the other defects. This can lead to a ``deadlock'' and terminate
cooperation. The client's long-term payoff can be given by
\begin{align}
 & \pi_{c}\left(\text{COOP},\text{COOP}\right)\nonumber \\
 & =du_{c}\left(\text{L},\text{H}\right)+\left(1-d\right)\left(u_{c}\left(\text{H},\text{H}\right)+w\pi_{c}\left(\text{COOP},\text{COOP}\right)\right).\label{eq:coopPayoffc1}
\end{align}
Rearranging \eqref{eq:coopPayoffc1} yields
\begin{align}
\pi_{c}\left(\text{COOP},\text{COOP}\right) & =\frac{\left(1-d\right)u_{c}\left(\text{H},\text{H}\right)+du_{c}\left(\text{L},\text{H}\right)}{1-\left(1-d\right)w}.\label{eq:coopPayoffc}
\end{align}
Similarly, the SP's long-term payoff equals
\begin{align}
\pi_{s}\left(\text{COOP},\text{COOP}\right) & =\frac{\left(1-d\right)u_{s}\left(\text{H},\text{H}\right)+du_{s}\left(\text{L},\text{H}\right)}{1-\left(1-d\right)w}.\label{eq:coopPayoffs}
\end{align}
On the one hand, the numerator is the one-round average payoff in
the fading channel, where $u_{c}\left(\text{H},\text{H}\right)$ and
$u_{s}\left(\text{H},\text{H}\right)$ are the average payoffs of
the client and the SP without channel outage. Note that channel fading
reduces the SP's average gain but may increase the client's gain if
$u_{c}\left(\text{H},\text{H}\right)<\left(1-w\right)u_{c}\left(\text{L},\text{H}\right)$.
On the other hand, the outage probability $d$ in the denominator
reflects the impact of packet loss on long-term cooperation. Even
if both parties adopt the COOP strategy, the cooperation may come
to a halt due to the channel fading, which severely undermines the
trust foundation. 

For players who always defect (denote this strategy as \emph{ALLD}),
they profit only in the first round against COOP. For the client,
\begin{align}
\pi_{c}\left(\text{ALLD},\text{COOP}\right) & =u_{c}\left(\text{D},\text{H}\right)=\Gamma\left(d\right)\tau.\label{eq:alldPayoffc}
\end{align}
For the SP, even if it intends to defect, the transaction of the client
may be lost in the first place:
\begin{align}
\pi_{s}\left(\text{COOP},\text{ALLD}\right) & =\left(1-d\right)u_{s}\left(\text{H},\text{D}\right)+du_{s}\left(\text{L},\text{H}\right)\nonumber \\
 & =\left(1-d\right)pt-c\tau.\label{eq:alldPayoffs}
\end{align}

More generally, we introduce the \emph{JDEF} strategy, where the SP
or the client defects in a certain round and attempts to restore cooperation
in the $j$-th round thereafter and repeat the same process later
on. For the client choosing not to pay for the service, the defection
will always succeed, as it is not influenced by channel uncertainty.
However, the attempt to recover cooperation in the $j$-th round could
fail if the transaction is lost, leading to the termination of cooperation
and blocking further gains. The long-term payoff for the client can
be given by
\begin{align}
 & \pi_{c}\left(\text{JDEF},\text{COOP}\right)\nonumber \\
 & =u_{c}\left(\text{D},\text{H}\right)+w^{j-1}\left(1-d\right)\left(u_{c}\left(\text{R},\text{H}\right)+w\pi_{c}\left(\text{JDEF},\text{COOP}\right)\right),
\end{align}
which can be rearranged as
\begin{align}
 & \pi_{c}\left(\text{JDEF},\text{COOP}\right)\nonumber \\
= & \frac{\left(1-d\right)\left(u_{c}\left(\text{D},\text{H}\right)+w^{j-1}u_{c}\left(\text{H},\text{R}\right)\right)+du_{c}\left(\text{L},\text{H}\right)}{1-\left(1-d\right)w^{j}},\label{eq:jroundLongC}
\end{align}
which uses the fact that $u_{c}\left(\text{D},\text{H}\right)=u_{c}\left(\text{L},\text{H}\right)$.
Similarly, for the SP, before the intentional defection, there is
a probability $d$ that the client's payment be lost, resulting in
a deadlock. If no loss occurs, the SP defects and then returns to
cooperation in the $j$-th round, the long-term payoff can be given
by
\begin{align}
 & \pi_{s}\left(\text{COOP},\text{JDEF}\right)=du_{s}\left(\text{L},\text{H}\right)+\left(1-d\right)u_{s}\left(\text{H},\text{D}\right)\nonumber \\
 & +\left(1-d\right)w^{j-1}\left(u_{s}\left(\text{H},\text{R}\right)+w\pi_{s}\left(\text{COOP},\text{JDEF}\right)\right).\label{eq:jroundLongSP1}
\end{align}
Rearranging \eqref{eq:jroundLongSP1}, we can derive that 
\begin{align}
 & \pi_{s}\left(\text{COOP},\text{JDEF}\right)=\nonumber \\
 & \frac{\left(1-d\right)\left(u_{s}\left(\text{H},\text{D}\right)+w^{j-1}u_{s}\left(\text{H},\text{R}\right)\right)+du_{s}\left(\text{L},\text{H}\right)}{1-\left(1-d\right)w^{j}}.\label{eq:jroundLongSP}
\end{align}
In \eqref{eq:jroundLongC} and \eqref{eq:jroundLongSP}, the numerator
represents the average payoff over the $j$ rounds. It is evident
that a higher packet loss rate decreases the average payoff for both
the client and the SP within these rounds. The term $\left(1-d\right)$
in the denominator integrates packet loss into the discounting process
for future benefits, highlighting that both parties must account both
for the subjective cooperation willingness $w$ and the objective
channel outage $d$. When $j=1$, the payoff of JDEF equals the payoff
of COOP. When $j=\infty$, it corresponds to the ALLD strategy, where
participants consistently defect each other. 

If a player randomly chooses to cooperate or defect with a certain
probability, i.e., a mixed strategy, \cite{cao2024} has shown that
the corresponding payoff cannot exceed both COOP and ALLD. Therefore,
we do not need to involve any mixed strategies in the following analysis.\vspace{-0.3cm}

\section{Cooperation Conditions under Channel Fading \label{sec:coopCond}}

\vspace{-0.3cm}In this section, we discuss the conditions for cooperation.
We first give the definition of the cooperation conditions.
\begin{defn}
\label{thm:CoopCondDef}(Cooperation conditions) Cooperation conditions
imply no strategy involving defection can obtain more payoff than
COOP when faced with COOP, i.e., the strategy profile (COOP, COOP)
is dominant.
\end{defn}
Definition \ref{thm:CoopCondDef} is the basis of the subsequent analysis.
As shown in Section \ref{subsec:strategyAnalysis}, JDEF is the most
general defective strategies where the SP or the client defects in
a round and then restores cooperation in a later round. Hence, as
long as the payoff of the strategy profile (COOP, COOP) exceeds that
of (JDEF, COOP) for any $j>1$, we can make sure that no strategy
involving defection can obtain more profits against COOP. We derive
the cooperation conditions in the following theorem.
\begin{thm}
\label{thm:nashCondw}The cooperation conditions are given by:
\begin{align}
w & \geq\frac{1-\tau/t}{\left(1-d\right)p/c-\tau/t},\label{eq:nashCondSPw}\\
w & \geq1-\frac{1-p/\Gamma\left(d\right)}{\tau/t}.\label{eq:nashCondCw}
\end{align}
\end{thm}
\begin{proof}
For the client, from the condition $\pi_{c}\left(\text{COOP},\text{COOP}\right)\geq\pi_{c}\left(\text{JDEF},\text{COOP}\right)$,
we have
\[
\left(1-d\right)\left(w^{j-1}-1\right)\left(\Gamma\left(d\right)\left(t-\left(1-w\right)\tau\right)-pt\right)\leq0,
\]
which can be rearranged as \eqref{eq:nashCondCw} since $d<1$ and
$w<1$. Similarly, for the SP, we can obtain \eqref{eq:nashCondSPw}
from $\pi_{c}\left(\text{COOP},\text{COOP}\right)\geq\pi_{s}\left(\text{COOP},\text{JDEF}\right)$. 
\end{proof}
Theorem \ref{thm:nashCondw} points out the conditions under which
cooperation can be achieved. The cooperation conditions in Theorem
\ref{thm:nashCondw} give a lower bound to cooperation willingness
$w$. As $d$ increases, the lower bound of $w$ also rises, indicating
that a higher outage probability, i.e., a poorer channel, requires
a stronger willingness to cooperate. Theorem \ref{thm:nashCondw}
also illustrates the impact of physical channel conditions. For the
client, $\Gamma\left(d\right)$ represents the impact of $d$ on the
client's utility, where an increase in $d$ results in decreased benefits
for the client. Similarly, for the SP, a larger outage probability
$d$ requires a high service price $p$ to maintain the cooperation.
This is because the client's payment may be lost in the fading channel,
thereby diminishing the revenue of the SP. 

Notably, by ensuring the payoff of (COOP, COOP) exceeds that of (ALLD,
COOP), we obtain the same conclusions as in Theorem \ref{thm:nashCondw}.
When faced with COOP, if the payoff of ALLD surpasses that of COOP,
then for any $j>1$, the payoff of JDEF will also exceed that of COOP.
This suggests that ALLD dominates all the strategies that include
defection. Therefore, once a participant defects in a certain round,
it will not restore cooperation in the best interest.

According to Theorem \ref{thm:nashCondw}, we can obtain the following
corollary in terms of channel outage.
\begin{cor}
\label{col:nashCondd}To guarantee cooperation, the channel outage
probability must be upper bounded by
\begin{align}
d\leq d_{s}\left(w;\tau,p\right) & \triangleq1-\frac{c\left(1-\left(1-w\right)\frac{\tau}{t}\right)}{wp},\label{eq:nashThreSPd}\\
d\leq d_{c}\left(w;\tau,p\right) & \triangleq\Gamma^{-1}\left(\frac{p}{1-\left(1-w\right)\frac{\tau}{t}}\right).\label{eq:nashThreCd}
\end{align}
\end{cor}
\begin{proof}
Rearranging \eqref{eq:nashCondSPw} and \eqref{eq:nashCondCw} yields
\eqref{eq:nashThreSPd} and \eqref{eq:nashThreCd}.
\end{proof}
Corollary \ref{col:nashCondd} implies the tolerance for channel outage
to establish mutual trust. \eqref{eq:nashThreSPd} and \eqref{eq:nashThreCd}
represent the outage thresholds below which the SP and the client
are inclined to cooperate, i.e., trust each other, respectively.

For the SP, it can be proved that $d_{s}\left(w;\tau,p\right)$ is
monotonically increasing in $w$, $\tau$, and $p$. On the one hand,
under a fixed $w$, increasing $\tau$ and $p$ can raise the SP's
cooperation threshold, encouraging cooperation over a wider range
of $d$. On the other hand, with fixed system configurations, a higher
$w$ increases the threshold, indicating that when the subjective
willingness to cooperate is strong, the SP exhibits greater tolerance
for channel uncertainty. Similarly, \eqref{eq:nashThreCd} sets the
client's cooperation threshold for $d$. It can be observed that $d_{c}\left(w;\tau,p\right)$
monotonically increases with $w$, but decreases with increasing $\tau$
and $p$. Thereby, reducing $\tau$ or $p$ raises the client's outage
threshold with a fixed $w$. Under specific system configurations,
a higher $w$ enables the client to adopt a cooperative strategy over
a wider range of $d$. 

To establish a cooperative relationship, the outage probability must
satisfy both thresholds:
\begin{equation}
d\leq d_{\text{min}}\left(w;\tau,p\right)\triangleq\min\left\{ d_{s}\left(w;\tau,p\right),d_{c}\left(w;\tau,p\right)\right\} ,\label{eq:dthre}
\end{equation}
where $d_{\text{min}}\left(w;\tau,p\right)$ is defined as the minimum
outage threshold for cooperation. If \eqref{eq:dthre} is not met,
an imbalance may arise where one party trusts and the other opts for
deception, severely compromising the trust foundation. The opposing
monotonicity of $d_{s}\left(w;\tau,p\right)$ and $d_{c}\left(w;\tau,p\right)$
for $\tau$ and $p$ implies that the outage thresholds mutually constrain
each other, which calls for a proper framework configuration to balance
both thresholds. 

Theorem \ref{thm:nashCondw} and Corollary \ref{col:nashCondd} indicate
that the cooperation requires both high commitment, i.e., a lower
bound on the cooperation willingness $w$, and high competence, i.e.,
an upper bound on the channel outage $d$. It is consistent with the
insight in \cite{Ling2025} that trustworthiness can be compromised
by either low commitment or low competence.\textcolor{blue}{}

Furthermore, we can derive a necessary condition for cooperation.
\begin{cor}
\label{thm:neceConddw}A necessary cooperation condition of the relationship
between $w$ and $d$ is
\begin{equation}
w>\frac{c}{\left(1-d\right)\Gamma\left(d\right)}.\label{eq:neceConddw}
\end{equation}
\end{cor}
\begin{proof}
Rearranging \eqref{eq:nashCondSPw} and \eqref{eq:nashCondCw} to
isolate $\tau$, we have
\begin{equation}
\frac{c-w\left(1-d\right)p}{\left(1-w\right)c}t<\tau<\frac{1-\nicefrac{p}{\Gamma\left(d\right)}}{1-w}.\label{eq:tauBound}
\end{equation}
Since the lower bound is lower than the upper bound,  \eqref{eq:neceConddw}
can be obtained. 
\end{proof}
Corollary \ref{thm:neceConddw} reveals the relationship between $w$
and $d$. Since $\Gamma\left(d\right)$ is monotonically decreasing
and $\Gamma\left(d\right)>0,0<d<1$, $\frac{c}{\left(1-d\right)\Gamma\left(d\right)}$
is also monotonically decreasing. Therefore, \eqref{eq:neceConddw}
indicates that, in a harsh wireless environment, $w$ is required
to be sufficiently large to achieve cooperation. Conversely, as $w$
decreases, a lower $d$ is required to achieve cooperation, implying
a higher channel quality requirement.
\begin{cor}
\label{thm:necessaryCond}A necessary cooperation condition for $w$
and $d$ is given by
\begin{align}
d & <\min\left\{ 1-\frac{c}{p},\Gamma^{-1}\left(p\right)\right\} ,\label{eq:neceCondd}\\
w & >\max\left\{ \frac{1-\tau/t}{p/c-\tau/t},1-\frac{1-p/\Gamma\left(0\right)}{\tau/t}\right\} .\label{eq:neceCondw}
\end{align}
\end{cor}
\begin{proof}
Since $w<1$, we can obtain from \eqref{eq:nashCondSPw} and \eqref{eq:nashCondCw}
that $d<1-\frac{c}{p}$ and $\Gamma\left(d\right)>p.$ As $\Gamma$
is monotonically decreasing, its inverse function must exist, yielding
\eqref{eq:neceCondd}. Similarly, combining $d>0$ with \eqref{eq:nashThreSPd}
and \eqref{eq:nashThreCd} gives \eqref{eq:neceCondw}.
\end{proof}
In Corollary \ref{thm:necessaryCond}, \eqref{eq:neceCondd} specifies
an upper bound of outage probability $d$ that must be satisfied to
build mutual trust in the system. Since $c$ and $\Gamma$ are determined
by the service itself, it is crucial to set the price $p$ to minimize
the ratio $c/p$ and maximize the value of $\Gamma^{-1}\left(p\right)$
so that cooperation can be effectively established over a poor channel
with a larger outage probability $d$.\vspace{-0.28cm}

\section{Optimization and Design\label{sec:SystemConfOpt}}

\subsection{Minimum Requirement on cooperation willingness\label{subsec:optw}}

\begin{figure}
\centering{}\begin{raggedright} \centering \includegraphics[width=0.45\textwidth]{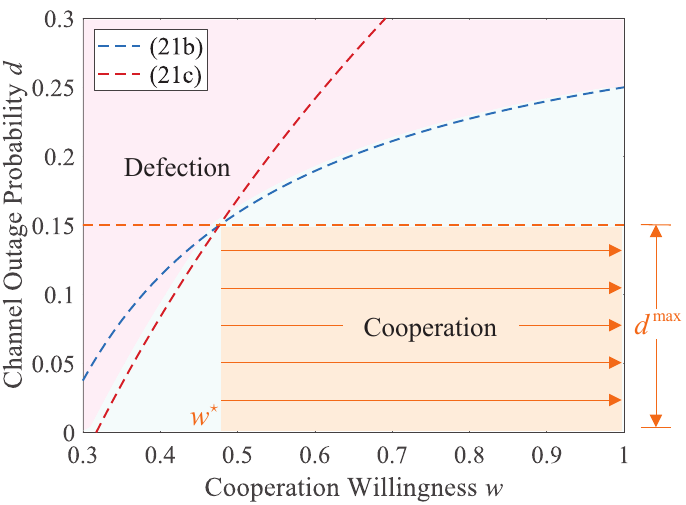}\end{raggedright}
\centering{}\caption{Minimum requirement on the cooperation willingness $w$, with $\Gamma\left(d\right)=\Gamma\left(0\right)\exp\left(-\varphi d\right)$
and $\varphi=2$, for $d^{\text{max}}=0.15$.\label{fig:minw}}
\vspace{-0.6cm}
\end{figure}
Theorem \ref{thm:nashCondw} points out the lower bound of $w$, i.e.,
the minimum cooperation willingness, to guarantee the service process.
In this subsection, we would like to minimize the requirement on
$w$ by optimizing system configurations $\tau$ and $p$, given the
worst-case channel condition $d\leq d^{\mathrm{max}}$, so the cooperation
can be established even when the willingness to cooperate is not strong
enough.

Mathematically, we formulate the above problem $\mathbb{P}_{w}^{\Gamma}\left(d^{\mathrm{max}}\right)$
as follows:\vspace{-0.1cm}\begin{subequations} 
\begin{align}
\mathbb{P}_{w}^{\Gamma}\left(d^{\mathrm{max}}\right):\;\underset{\tau,p}{\text{minimize}}\quad & w\label{fm:objw}\\
\text{subject to}\quad & w>\frac{1-\tau/t}{\left(1-d\right)p/c-\tau/t},\nonumber \\
 & \quad\quad\quad\quad\quad\forall0<d\leq d^{\text{max}},\label{fm:coopCondws}\\
 & w>1-\frac{1-p/\Gamma\left(d\right)}{\tau/t},\nonumber \\
 & \quad\quad\quad\quad\quad\forall0<d\leq d^{\text{max}},\label{fm:coopCondwc}\\
 & c<p<\Gamma\left(d\right),\enspace0<\tau<t,\label{fm:sysConf}
\end{align}
\end{subequations}where \eqref{fm:coopCondws} and \eqref{fm:coopCondwc}
represent the cooperation conditions on the SP's and the client's
sides according to Theorem \ref{thm:nashCondw}, respectively. These
two conditions must be satisfied for any $d\leq d^{\text{max}}$.
The constraint in \eqref{fm:sysConf} specifies the feasible region
for the system configuration $p$ and $\tau$. Highlight that $\mathbb{P_{\mathrm{w}}^{\mathrm{\Gamma}}}\left(d^{\text{max}}\right)$
is a nonconvex problem due to the non-convexity of the constraints
\eqref{fm:coopCondws} and \eqref{fm:coopCondwc}. 

To solve $\mathbb{P}_{w}^{\Gamma}\left(d^{\text{max}}\right)$, we
use a two-stage optimization process. First, we optimize $\tau$ and
$p$ to obtain the minimum $w$ that guarantees cooperation for a
given $d$, expressed as\vspace{-0.1cm}
\begin{align*}
\mathbb{P}_{w,1}^{\Gamma}\left(d\right):\;\underset{\tau,p}{\text{minimize}}\quad & w\\
\text{subject to}\quad & w>\frac{1-\tau/t}{\left(1-d\right)p/c-\tau/t},\\
 & w>1-\frac{1-p/\Gamma\left(d\right)}{\tau/t},\\
 & c<p<\Gamma\left(d\right),0<\tau<t.
\end{align*}
One can see that $\frac{1-\tau/t}{\left(1-d\right)p/c-\tau/t}$ is
monotonically decreasing in both $\tau$ and $p$, and $1-\frac{1-p/\Gamma\left(d\right)}{\tau/t}$
is monotonically increasing in them. The minimum $w^{\star}\left(d\right)$
can be determined by letting them equal to each other, i.e.,
\begin{equation}
w^{\star}\left(d\right)=\frac{1-\tau/t}{\left(1-d\right)p/c-\tau/t}=1-\frac{1-p/\Gamma\left(d\right)}{\tau/t}.\label{eq:optw11}
\end{equation}
Rearranging \eqref{eq:optw11}, we can obtain that
\begin{equation}
w^{\star}\left(d\right)=\frac{c}{\left(1-d\right)\Gamma\left(d\right)},\label{eq:optw1}
\end{equation}
where $p$ and $\text{\ensuremath{\tau}}$ must satisfy the following
relationship:
\begin{equation}
p=\Gamma\left(d\right)\left(1-\frac{\text{\ensuremath{\tau}}}{t}\right)+\frac{c}{\left(1-d\right)}\frac{\text{\ensuremath{\tau}}}{t}.\label{eq:optw_ptau1}
\end{equation}
As one can see, $w^{\star}\left(d\right)$ represents the minimum
cooperation willingness required for trust establishment under the
specified channel outage probability $d$. For a given $d$, as long
as $\tau$ and $p$ are set according to \eqref{eq:optw_ptau1}, the
cooperation between the SP and the client can be maintained with the
lowest possible cooperation willingness $w^{\star}\left(d\right)$. 

In the second stage, we maximize the minimum requirement on cooperation
willingness $w^{\star}\left(d\right)$ to ensure the cooperation under
any channel conditions $d\leq d^{\mathrm{max}}$, which is formulated
as
\begin{align*}
\mathbb{P}_{w,2}^{\Gamma}\left(d^{\mathrm{max}}\right):\;\underset{d}{\text{maximize}}\quad & w^{\star}\left(d\right)\\
\text{subject to}\quad & 0<d\leq d^{\mathrm{max}}.
\end{align*}
We can easily prove that $w^{\star}\left(d\right)$ in \eqref{eq:optw1}
is monotonically increasing by calculating its first-order derivative
(since $\Gamma^{\prime}\left(x\right)<0$ and $\Gamma\left(x\right)>0$).
Therefore, $w^{\star}\left(d\right)$ is maximized at $d=d^{\mathrm{max}}$.
As a result, the solution to $\mathbb{P}_{w}^{\Gamma}\left(d^{\mathrm{max}}\right)$
is given by
\begin{align}
p^{\star} & =\Gamma\left(d^{\text{max}}\right)\left(1-\frac{\text{\ensuremath{\tau^{\star}}}}{t}\right)+\frac{c}{\left(1-d^{\text{max}}\right)}\frac{\text{\ensuremath{\tau^{\star}}}}{t},\label{eq:optw_ptau}\\
 & \quad\forall\tau^{\star}\in\left(0,\frac{\Gamma\left(d^{\text{max}}\right)-c}{\Gamma\left(d^{\text{max}}\right)-c/\left(1-d^{\text{max}}\right)}t\right),\label{eq:optw_taurg}
\end{align}
and the corresponding minimum requirement on cooperation willingness
$w$ is
\begin{equation}
w^{\star}=\frac{c}{\left(1-d^{\text{max}}\right)\Gamma\left(d^{\text{max}}\right)}.\label{eq:optw}
\end{equation}
The above results indicate that, by setting $\tau$ and $p$ according
to \eqref{eq:optw_ptau}, we can minimize the requirement on the cooperation
willingness $w$ to guarantee cooperation under the worst outage probability
$d^{\text{max}}$. That is, the interactive wireless assess services
can be established over any channel satisfying $d<d^{\mathrm{max}}$,
as long as the cooperation willingness of both sides is larger than
$\frac{c}{\left(1-d^{\text{max}}\right)\Gamma\left(d^{\text{max}}\right)}$.
Also, \eqref{eq:optw} shows that a lower $d^{\mathrm{max}}$ reduces
the minimum requirement on cooperation willingness $w^{\star}$, making
the cooperation easier to be established. The SP can increase transmission
power or improve the channel quality, i.e., a smaller worse-case outage
$d^{\mathrm{max}}$, to lower the requirement on $w^{\star}$ and
facilitate the long-term cooperation. 

Combining \eqref{eq:optw_ptau} and \eqref{eq:optw_taurg} yields
$c/\left(1-d^{\text{max}}\right)<p^{\star}<\Gamma\left(d^{\text{max}}\right)$,
indicating that a larger $d^{\text{max}}$ reduces the possible range
of the price $p$. On the one hand, it is because a poorer channel
harms the client's utility and reduces the maximum price that the
client can accept; on the other hand, a higher outage probability
increases the risk of payment losses, thus requiring a higher price
from the SP's perspective. Consequently, as $d^{\text{max}}$ increases,
the payoffs of both parties are affected, and thus the feasible pricing
range narrows. 

Fig. \ref{fig:minw} illustrates an example with $c=0.3$, $\Gamma\left(0\right)=1$,
and $d^{\text{max}}=0.15$. We use the utility function $\Gamma\left(d\right)=\Gamma\left(0\right)\exp\left(-\varphi d\right)$
with $\varphi=2$. Fig. \ref{fig:minw} shows that, for $\tau$ and
$p$ satisfying \eqref{eq:optw_ptau}, the constraints in \eqref{fm:coopCondws}
and \eqref{fm:coopCondwc} for the SP and the client intersect at
$d=d^{\text{max}}$. Hence, as long as the cooperation willingness
exceeds $w^{\star}$ in \eqref{eq:optw}, the cooperation can be guaranteed
over any channels satisfying $d<d^{\text{max}}$ (see the orange area.)

\begin{cor}
\label{thm:necesuff}The necessary condition $w>\frac{c}{\left(1-d\right)\Gamma\left(d\right)}$
in Corollary \ref{thm:neceConddw} is also sufficient, if $p$ and
$\tau$ satisfy \eqref{eq:optw_ptau1}.
\end{cor}
\begin{proof}
If $p$ and $\tau$ satisfy \eqref{eq:optw_ptau1}, the cooperation
conditions in \eqref{eq:nashCondSPw} and \eqref{eq:nashCondCw} is
equivalent to $w\geq\frac{c}{\left(1-d\right)\Gamma\left(d\right)}$. 
\end{proof}
Corollary \ref{thm:necesuff} indicates that, if $p$ and $\tau$
are properly set, the necessary condition in Corollary \ref{thm:neceConddw}
is also sufficient, which holds for both the client and the SP. Instead
of the original \textbf{step2} in Fig. \ref{fig:workflow} where the
SP and client need negotiate service terms with proper parameters,
Corollary \ref{thm:necesuff} provides another low-complex approach
to achieve the cooperation: 1) The SP measures the channel outage
$d$ and sets the parameters $p$ and $\tau$ according to \eqref{eq:optw_ptau1};
2) Both parties assess the cooperation condition in \eqref{eq:neceConddw}
before the service. With the help of Corollary \ref{thm:necesuff},
the parameters can be determined and set in a simple way.

\vspace{-0.3cm}

\subsection{Minimum Requirement on Channel Outage \label{subsec:optd}}

\begin{figure}
\centering{}\begin{raggedright} \centering \includegraphics[width=0.45\textwidth]{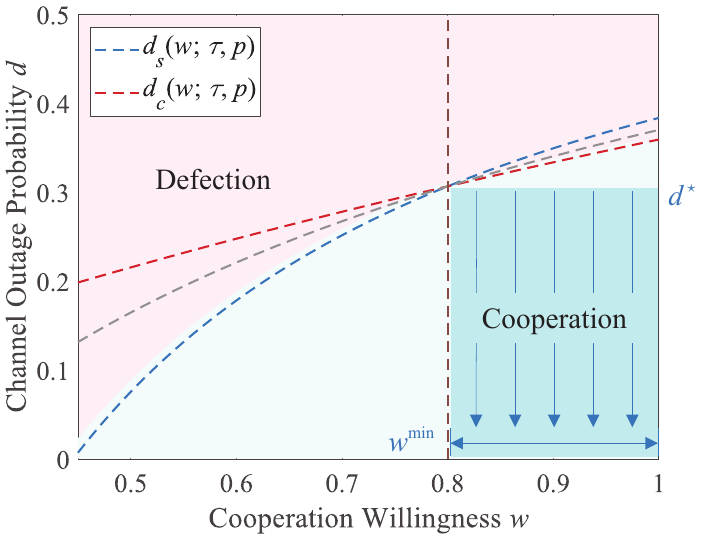}\end{raggedright}
\centering{}\caption{Minimum requirement on the channel outage probability $d$, with $\Gamma\left(d\right)=\Gamma\left(0\right)\exp\left(-\varphi d\right)$
and $\varphi=2$, for $w^{\text{min}}=0.8$. \label{fig:maxd}}
\vspace{-0.6cm}
\end{figure}
As shown in Section \ref{sec:coopCond}, a harsher channel challenges
the trust establishment between the client and the SP. In this subsection,
we aim to minimize the requirement on the channel quality $d$, given
the least willingness to cooperate, i.e., $w\geq w^{\text{min}}$.
Similar to $\mathbb{P}_{w}^{\Gamma}\left(d^{\text{max}}\right)$,
we formulate the optimization problem $\mathbb{P}_{d}^{\Gamma}\left(w^{\text{min}}\right)$
as:
\begin{align*}
\mathbb{P}_{d}^{\Gamma}\left(w^{\text{min}}\right):\;\underset{\tau,p}{\text{maximize}}\quad & d\\
\text{subject to}\quad & d\leq d_{s}\left(w;\tau,p\right),\forall w^{\text{min}}\leq w\leq1,\\
 & d\leq d_{c}\left(w;\tau,p\right),\forall w^{\text{min}}\leq w\leq1,\\
 & c<p<\Gamma\left(d\right),0<\tau<t.
\end{align*}
We use the same two-stage optimization method in Section \ref{subsec:optw},
by optimizing $\tau$ and $p$ first and then $w$. The worst channel
condition $d^{\star}$ to guarantee the cooperation is obtained at
$w=w^{\text{min}}$, which is given by:
\begin{align}
d^{\star} & =\arg_{x}\ f\left(x\right)=c/w^{\text{min}},\label{eq:optd}
\end{align}
where $f\left(x\right)=\left(1-x\right)\Gamma\left(x\right)$. One
can see that $f\left(x\right)$ is monotonically increasing within
$0<x<1$. Since $f\left(0\right)=\Gamma\left(0\right)>c/w^{\text{min}}$
(from \eqref{eq:neceConddw}) and $f\left(1\right)=0<c/w^{\text{min}}$,
there exists a unique root $d^{\star}$ satisfying $f\left(x\right)=c/w^{\text{min}}$.
The solution to $\mathbb{P_{\mathrm{d}}^{\mathrm{\Gamma}}}\left(w^{\text{min}}\right)$
is given by
\begin{align}
p & =\Gamma\left(d^{\star}\right)\left(1-\left(1-w^{\text{min}}\right)\frac{\text{\ensuremath{\tau}}}{t}\right),\label{eq:optd_ptau}\\
 & \forall\tau\in\left(0,\frac{w^{\text{min}}-\left(1-d^{\star}\right)c}{w^{\text{min}}-c}t\right).\label{eq:optd_taurg}
\end{align}
The above optimization minimizes the requirement on channel outage
to guarantee the cooperation. By setting $\tau$ and $p$ to satisfy
\eqref{eq:optd_ptau} and \eqref{eq:optd_taurg}, cooperation is achieved
for $w>w^{\text{min}}$, if the outage probability is lower than $d^{\star}$.
When $w^{\text{min}}$ is higher, the requirement on $d^{\star}$
also increases, indicating that a higher cooperation willingness helps
establish trust under harsh environment. A shorter slot time $t$
can improve the worst-case cooperation willingness $w^{\text{min}}$
and lower the requirement on channel outage. However, it also increases
the number of rounds to complete a service and raises the risk of
payment losses.

Fig. \ref{fig:maxd} presents an example with $c=0.3$, $\Gamma\left(0\right)=1$,
and $w^{\text{min}}=0.1$ and uses the utility function $\Gamma\left(d\right)=\Gamma\left(0\right)\exp\left(-\varphi d\right)$
with $\varphi=2$. The trial time $\tau$ and service price $p$ are
set according to \eqref{eq:optd_ptau}. One can observe that $d_{s}\left(w;\tau,p\right)$
and $d_{c}\left(w;\tau,p\right)$ intersect at $w=w^{\text{min}}$,
and the cooperation can be guaranteed for $w>w^{\text{min}}$, as
long as the channel outage probability is below $d^{\star}$.

\vspace{-0.3cm}

\subsection{Joint Optimization}

In the previous two subsections, we optimize the framework by considering
either the least cooperation willingness $w^{\text{min}}$ or the
worst channel condition $d^{\text{max}}$.  However, these two factors
may exist simultaneously in practice. Hence, we would like to design
the system, given both the worst-case cooperation willingness $w^{\text{min}}$
and channel outage $d^{\text{max}}$. Interestingly, we find that
the above two requirements can be satisfied at the same time, which
is summarized by the following theorem. \textcolor{magenta}{}\textcolor{blue}{}
\begin{thm}
\label{thm:jointOpt} $\mathbb{P}_{d}^{\Gamma}\left(w^{\text{min}}\right)$
and $\mathbb{P}_{w}^{\Gamma}\left(d^{\text{max}}\right)$ can be optimized
simultaneously. The optimal solution $\tau^{\star}$ and $p^{\star}$
are given by 
\begin{align}
p^{\star} & =\frac{c\left(\frac{c}{w^{\text{min}}}-\Gamma\left(d^{\text{max}}\right)\left(1-d^{\text{max}}\right)\right)}{\left(\Gamma\left(d^{\star}\right)-\Gamma\left(d^{\text{max}}\right)\right)\left(1-d^{\star}\right)\left(1-d^{\text{max}}\right)-c\left(d^{\star}-d^{\text{max}}\right)},\label{eq:jointdw_p}\\
\tau^{\star} & =\frac{\Gamma\left(d^{\text{max}}\right)-p^{\star}}{\Gamma\left(d^{\text{max}}\right)-c/\left(1-d^{\text{max}}\right)}t,\label{eq:jointdw_tau}
\end{align}
where $d^{\star}=\arg_{x}\ f\left(x\right)=c/w^{\text{min}}.$
\end{thm}
\begin{proof}
If $p$ and $\tau$ satisfy \eqref{eq:optw_ptau}, one can obtain
the minimum requirement on $w$ for any $d\leq d^{\text{max}}$. Similarly,
if they satisfy \eqref{eq:optd_ptau}, we obtain the necessary requirement
on $d$ for any $w\geq w^{\text{min}}$. When both \eqref{eq:optw_ptau}
and \eqref{eq:optd_ptau} are simultaneously satisfied, the above
two requirements can be met at the same time. As a result, $\mathbb{P}_{d}^{\Gamma}\left(w^{\text{min}}\right)$
and $\mathbb{P}_{w}^{\Gamma}\left(d^{\text{max}}\right)$ are jointly
optimized, and $p^{\star}$ and $\tau^{\star}$ are given by \eqref{eq:jointdw_p}
and \eqref{eq:jointdw_tau}. 
\end{proof}
Theorem \ref{thm:jointOpt} indicates that $\mathbb{P}_{d}^{\Gamma}\left(w^{\text{min}}\right)$
and $\mathbb{P}_{w}^{\Gamma}\left(d^{\text{max}}\right)$ can be jointly
optimized under the worst channel conditions and least cooperation
willingness. The minimum requirements on $d$ and $w$ are given by
$d^{\star}=\arg_{x}\ f\left(x\right)=c/w^{\text{min}}$ and $w^{\star}=\frac{c}{\left(1-d^{\text{max}}\right)\Gamma\left(d^{\text{max}}\right)}$,
as long as the channel uncertainty and cooperation willingness satisfy
$d\leq d^{\text{max}}$ and $w\geq w^{\text{min}}$. Theorem \ref{thm:jointOpt}
lowers the thresholds for cooperations on $d$ and $w$ simultaneously
and thus enhances the robustness and stability of the system. 

\begin{figure}
\centering{}\begin{raggedright} \centering \includegraphics[width=0.45\textwidth]{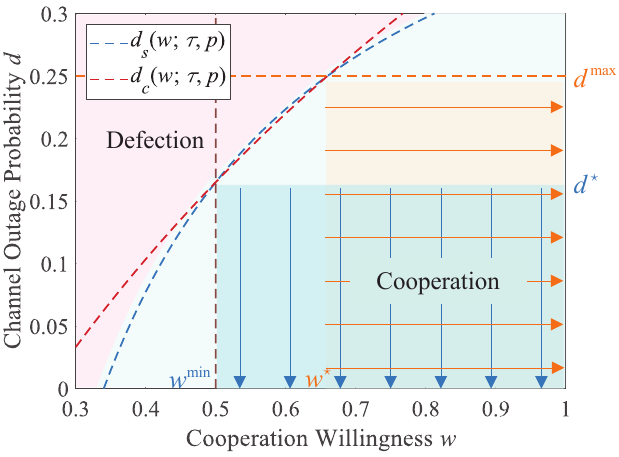}\end{raggedright}
\centering{}\caption{Joint optimization for channel outage probability $d$ and cooperation
willingness $w$, with $\Gamma\left(d\right)=\Gamma\left(0\right)\exp\left(-\varphi d\right)$
and $\varphi=2$ for $d^{\text{max}}=0.25$ and $w^{\text{min}}=0.5$.\label{fig:jointdw}}
\vspace{-0.6cm}
\end{figure}
In Fig. \ref{fig:jointdw}, the parameters are set as $c=0.3$, $\Gamma\left(0\right)=1$,
$w^{\text{min}}=0.55$, and $d^{\text{max}}=0.2$. We use the utility
function $\Gamma\left(d\right)=\Gamma\left(0\right)\exp\left(-\varphi d\right)$
with $\varphi=2$. When $p$ and $\tau$ satisfy \eqref{eq:jointdw_p}
and \eqref{eq:jointdw_tau}, $d_{s}\left(w;\tau,p\right)$ and $d_{c}\left(w;\tau,p\right)$
intersect at $w=w^{\mathrm{min}}$ and $d=d^{\text{max}}$ simultaneously.
Hence, the requirements on both $d$ and $w$ are minimized, and the
cooperation can be ensured if the above requirements are satisfied.\textcolor{red}{{}
}\vspace{-0.3cm}

\section{Cooperation Area \label{sec:CoopArea}}

\subsection{Definition}

\begin{figure*}
\begin{raggedright} \centering\includegraphics[width=1\textwidth]{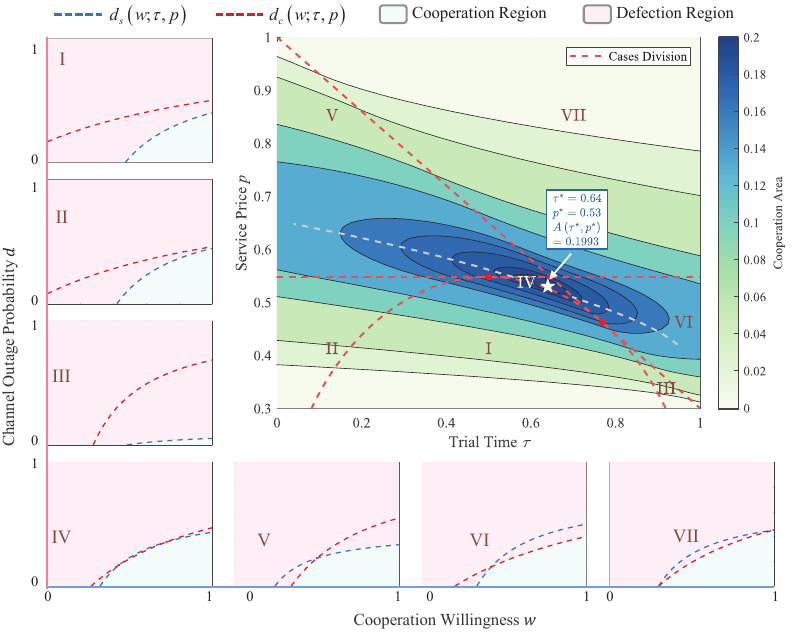}\end{raggedright}
\centering{}

\caption{Illustration of all the seven cases in analyzing the cooperation region.\textcolor{blue}{\label{fig:coopModes}}}
\vspace{-0.6cm}
\end{figure*}
In Section \ref{sec:SystemConfOpt}, we consider the framework optimization
under some worst-case assumptions. However, this approach is still
limited, as it merely considers specific worst-case conditions rather
than the overall system performance. To achieve cooperation across
more general scenarios, we would like to extend the analysis to a
more comprehensive optimization of $d$ and $w$. In this section,
we aim to maximize the feasible range of $d$ and $w$ that can guarantee
the cooperation.  To comprehensively assess the impact of the channel
quality $d$ and cooperation willingness $w$ on the trust-building
process, we introduce the concept of cooperation region, which is
defined as the set of all the possible $d$ and $w$ satisfying the
cooperation conditions:
\[
{\cal A}\left(\tau,p\right)=\left\{ \left(d,w\right)|d<d_{\min}\left(w;\tau,p\right),0<d,w<1\right\} .
\]
Note that $d_{\min}\left(w;\tau,p\right)$, defined in \eqref{eq:dthre},
is the minimum outage threshold for cooperation. Furthermore, we define
the cooperation area $A\left(\tau,p\right)$ as the area of cooperation
region:
\begin{align*}
A\left(\tau,p\right) & \triangleq\int_{{\cal A}\left(\tau,p\right)}\text{d}\left(d,w\right)\\
 & =\int_{0}^{1}\max\left(d_{\min}\left(w;\tau,p\right),0\right)\text{d}w.
\end{align*}

The cooperation area reflects the framework robustness to guarantee
the cooperation in uncertain environment. The value of the cooperation
area $A\left(\tau,p\right)$ is determined by the trial time $\tau$
and service price $p$. Therefore, to maximize the system robustness
under varying cooperation willingness and channel uncertainty, we
try to seek the optimal $p$ and $\tau$ that maximize the cooperation
area $A\left(\tau,p\right)$, which can be formulated as follows:\begin{subequations}
\begin{align}
\mathbb{P}_{A}^{\Gamma}:\;\underset{\tau,p}{\text{maximize}}\quad & A\left(\tau,p\right)\label{fm:coopArea}\\
\text{subject to}\quad & c<p<\Gamma\left(0\right),0<\tau<t.\label{fm:stptau}
\end{align}
\end{subequations} The above problem is quite challenging and more
difficult than the design problems in Section \ref{sec:SystemConfOpt}
due to the complicated expression of $A\left(\tau,p\right)$. We even
can hardly express the objective function in a closed form, not to
mention its properties.\textcolor{blue}{{} }In the following analysis,
we will adopt the linear utility function $\Gamma\left(d\right)=\Gamma\left(0\right)\left(1-d\right)$.
\vspace{-0.3cm}

\subsection{Divide and Characterize \label{subsec:coopAreaCha}}

Note that, even if a linear utility function is used, the cooperation
area maximization is still challenging. $A\left(\tau,p\right)$ does
not have a unified expression, making it difficult to derive an optimal
solution. Hence, we must divide the cooperation region as follows.
\begin{prop}
\label{thm:7cases}The cooperation area $A\left(\tau,p\right)$ can
be divided into seven cases, listed in Table \ref{tab:coopModes}.
\end{prop}
\begin{proof}
Please see Appendix \ref{subsec:ProofProposition1}.
\end{proof}
The seven cases are visualized in Fig. \ref{fig:coopModes} and mathematically
described in Table \ref{tab:coopModes}. ``Dominant'' in Table \ref{tab:coopModes}
represents the party that imposes a stricter constraint for channel
outage ($d_{s}\left(w;\tau,p\right)$ or $d_{c}\left(w;\tau,p\right)$).
For example, in Case II, when the outage probability is below $d_{c}\left(w;\tau,p\right)$,
the client trusts the SP and pays for the service on time. However,
it may be still higher than $d_{s}\left(w;\tau,p\right)$, and the
SP defects by not delivering the service after the payment. In this
case, the SP is dominated in the trust relationship, because such
a mismatch puts the client in a weaker position, significantly harming
its benefit and undermining mutual trust.\textcolor{blue}{{} }\textcolor{red}{}\textcolor{blue}{}

These seven cases are divided according to the intersection points
between $d_{s}\left(w;\tau,p\right)$ and $d_{c}\left(w;\tau,p\right)$,
since the expression of $A\left(\tau,p\right)$ highly depends on
these intersection points. Case I has no intersection. In Case II,
all the intersections occur at $w>1$, and in Case III they all fall
below $d<0$. In these three cases, the cooperation threshold of the
SP is consistently lower than that of the client, indicating that
the trust is entirely SP-dominated. In Case IV, intersections are
within $0<d,w<1$. In Cases V and VI, one intersection is within the
feasible range, while the other is outside.\textcolor{red}{{} }These
three cases indicate a relatively balanced trust relationship, where
the dominant party varies from different cooperation willingness or
channel conditions. Case VII is the only case where trust is primarily
influenced by the client, with one intersection occurring at $d<0$
and the other at $w>1$.\vspace{-0.5cm}

\subsection{Cooperation Area Maximization}

After dividing the cooperation region into seven cases, we would like
to seek the global maximum by solving each of them.\textcolor{blue}{{}
}Due to the complexity of $A\left(\tau,p\right)$, we can hardly identify
the optimal solutions of each case. Despite these challenges, our
analysis still provides valuable insights into how the dominance of
trust influences the potential for cooperation. 

We begin with the optimization in Cases II, III, and VII, where trust
is dominated by a single party. We can derive the following proposition.
\begin{prop}
\label{thm:optCase237}The cooperation areas in Cases II, III and
VII are maximized at $\left(\tau^{\star},p^{\star}\right)=\left(\frac{1}{2}t\text{,}\sqrt{c\Gamma\left(0\right)}\right)$,
$\left(\frac{\Gamma\left(0\right)}{\Gamma\left(0\right)+c}t,2\frac{\Gamma\left(0\right)c}{\Gamma\left(0\right)+c}\right)$
and $\left(\frac{\Gamma\left(0\right)-\sqrt{c\Gamma\left(0\right)}}{\Gamma\left(0\right)-c}t,\sqrt{c\Gamma\left(0\right)}\right)$,
respectively.
\end{prop}
\begin{proof}
Please see Appendix \ref{subsec:ProofProposition2}.
\end{proof}
From Proposition \ref{thm:optCase237}, for SP-dominated trust in
Case II or III, $\left(\tau^{\star},p^{\star}\right)=\left(\frac{1}{2}t\text{,}\sqrt{c\Gamma\left(0\right)}\right)$
or $\left(\frac{\Gamma\left(0\right)}{\Gamma\left(0\right)+c}t,2\frac{\Gamma\left(0\right)c}{\Gamma\left(0\right)+c}\right)$
can maximize the cooperation area. For client-dominated trust in Case
VII, the optimal setting is $\left(\frac{\Gamma\left(0\right)-\sqrt{c\Gamma\left(0\right)}}{\Gamma\left(0\right)-c}t,\sqrt{c\Gamma\left(0\right)}\right)$.
Notably, these points fall on the boundary of Case IV, where trust
is jointly influenced by both parties. For Cases V and VI, we have
the following proposition.
\begin{prop}
\label{thm:optCase56}Define $A_{s/c}\left(\tau,p;x,y\right)\triangleq\int_{x}^{y}d_{s/c}\left(w;\tau,p\right)\text{d}w$.
The cooperation area of Case V is maximized, only if $p^{\star}$
and $\tau^{\star}$ satisfy 
\begin{equation}
A_{c}\left(\tau^{\star},p^{\star};w_{1},1\right)-A_{s}\left(\tau^{\star},p^{\star};w_{0},w_{1}\right)=1-2w_{1}+w_{0};\label{eq:optArea3}
\end{equation}
the cooperation area of Case VI is maximized, only if $p^{\star}$
and $\tau^{\star}$ satisfy 
\begin{equation}
A_{s}\left(\tau^{\star},p^{\star};w_{2},1\right)-A_{c}\left(\tau^{\star},p^{\star};w_{0},w_{2}\right)=1-2w_{2}+w_{0},\label{eq:optArea7}
\end{equation}
where 
\begin{align*}
w_{0} & =\max\left\{ \frac{c\left(t-\tau\right)}{pt-c\tau},\frac{pt-\Gamma\left(0\right)\left(t-\tau\right)}{\Gamma\left(0\right)\tau}\right\} ,\\
w_{1} & =\left(\frac{p-\sqrt{p^{2}-4c\Gamma\left(0\right)\frac{\tau}{t}\left(1-\frac{\tau}{t}\right)}}{2\frac{\tau}{t}\sqrt{c\Gamma\left(0\right)}}\right)^{2},\\
w_{2} & =\left(\frac{p+\sqrt{p^{2}-4c\Gamma\left(0\right)\frac{\tau}{t}\left(1-\frac{\tau}{t}\right)}}{2\frac{\tau}{t}\sqrt{c\Gamma\left(0\right)}}\right)^{2}.
\end{align*}
\end{prop}
\begin{proof}
Please see Appendix \ref{subsec:ProofProposition3}.
\end{proof}
By combining Propositions \ref{thm:optCase237} and \ref{thm:optCase56},
we can derive the following theorem.
\begin{table*}
\caption{The mathematical description of the seven cases of cooperation region
in terms of $\tau$ and $p$.\label{tab:coopModes}}

\begin{raggedright} \centering\renewcommand{\arraystretch}{1.45}%
\begin{tabular}{|c|c|c|c|c|c|c|}
\hline 
Cases & \multicolumn{5}{c|}{Boundaries} & Dominant\tabularnewline
\hline 
\hline 
I & \multirow{7}{*}{$p>c$} & \multicolumn{4}{c|}{$p^{2}<4c\Gamma\left(0\right)\frac{\tau}{t}\left(1-\frac{\tau}{t}\right)$} & SP\tabularnewline
\cline{1-1}\cline{3-7}
II &  & \multirow{6}{*}{$p^{2}\geq4c\Gamma\left(0\right)\frac{\tau}{t}\left(1-\frac{\tau}{t}\right)$} & \multirow{4}{*}{$p<\Gamma\left(0\right)\left(1-\frac{\tau}{t}\right)+c\frac{\tau}{t}$} & \multirow{3}{*}{$p<\sqrt{c\Gamma\left(0\right)}$} & $\tau<\frac{1}{2}t$ & SP\tabularnewline
\cline{1-1}\cline{6-7}
III &  &  &  &  & $\tau>\frac{\Gamma\left(0\right)}{\Gamma\left(0\right)+c}t$ & SP\tabularnewline
\cline{1-1}\cline{6-7}
IV &  &  &  &  & $\frac{1}{2}t<\tau<\frac{\Gamma\left(0\right)}{\Gamma\left(0\right)+c}t$ & Both\tabularnewline
\cline{1-1}\cline{5-7}
V &  &  &  & \multicolumn{2}{c|}{$p>\sqrt{c\Gamma\left(0\right)}$} & Both\tabularnewline
\cline{1-1}\cline{4-7}
VI &  &  & \multirow{2}{*}{$p>\Gamma\left(0\right)\left(1-\frac{\tau}{t}\right)+c\frac{\tau}{t}$} & \multicolumn{2}{c|}{$p<\sqrt{c\Gamma\left(0\right)}$} & Both\tabularnewline
\cline{1-1}\cline{5-7}
VII &  &  &  & \multicolumn{2}{c|}{$p>\sqrt{c\Gamma\left(0\right)}$} & Client\tabularnewline
\hline 
\end{tabular}\end{raggedright} \centering{}\vspace{-0.6cm}
\end{table*}

\begin{thm}
\label{thm:optArea}The cooperation area is maximized if $p$ and
$\tau$ are in Case IV, or in Case V satisfying \eqref{eq:optArea3},
or in Case VI satisfying \eqref{eq:optArea7}.
\end{thm}
\begin{proof}
We can prove that $A_{1}\left(\tau,p\right)$, the cooperation area
in Case I, is monotonically increasing in $p$. Therefore, its maximum
value must be on the boundary $p=\Gamma\left(0\right)\left(1-\frac{\tau}{t}\right)+c\frac{\tau}{t}$,
which is also the boundary of Cases II, III, and IV. Notably, the
local optima of Cases II, III and VII (pointed out by Proposition
\ref{thm:optCase237}) are at the boundary of Case IV. Therefore,
Cases I, II, III, and VII can be reduced to the boundary of Case IV.
The global optimal solution will be either in Case IV (and its boundary),
or in Case V satisfying \eqref{eq:optArea3}, or in Case VI satisfying
\eqref{eq:optArea7}.
\end{proof}
\vspace{-0.2cm}Theorem \ref{thm:optArea} shows that, the optimal
$p$ and $\tau$ to maximize the cooperation area must be in Cases
IV, V, or VI, where the trust is dominated by both parties. This implies
that in a zero-trust environment, it is problematic if the trust-building
process heavily relies on one party. A well-designed trust-building
framework should take both sides into consideration, and the participation
of both sides can enhance the fairness and balance of the entire cooperation
framework. Note that the closed-form solutions revealed in Proposition
\ref{thm:optCase237} also fall at the boundary of Case IV. Although
they are not globally optimal, they can be viewed as sub-optimal solutions
with acceptable performance in balancing trust relationship. 

Furthermore, we also find that the pricing strategy can affect the
trust relationship. For example, in Case VII, the service price $p$
is relatively high. The client is more likely to be dishonest since
the service is expensive. In this case, trust is dominated by the
client. Conversely, in Cases I, II, or III, the price $p$ is relatively
low. Although the client is willing to cooperate, the SP is more inclined
to defect, implying that trust is dominated by the SP. Hence, either
high or low $p$ leads to imbalanced trust relationships and reduces
the cooperation area.

Similarly, trust can also be influenced by the setting of the trial
time $\tau$. The trial time of Case VII is long, where the client
may have fulfilled most of their requirement during the trial service
and defect by not paying. Hence, trust is dominated by the client
in this case. Conversely, in Cases I, II, and III, the trial time
$\tau$ is short. The SP may reject to deliver the remainder services
which is much longer than the trial ones, since they have received
full payment. As a result, the SP is more likely to end the service
after the payment, and trust is dominated by the SP. In conclusion,
improper price or trial time makes the trust imbalanced and harms
the cooperation.

Fig. \ref{fig:coopModes} uses $c=0.3$ and $\Gamma\left(0\right)=1$
to visualize the cooperation area $A\left(\tau,p\right)$ for different
$p$ and $\tau$. The red dashed lines divide different seven cases
in Table \ref{tab:coopModes}. The red dots mark the local optimal
points in Corollary \ref{thm:optCase237}, while the grey dashed lines
represent the possible optimal solutions in Cases V and VI in Theorem
\ref{thm:optCase56}. The white star marker represents the global
optimum of the cooperation area (obtained via exhaustive search),
which verifies Theorem \ref{thm:optArea} as it appears in Case IV.
Fig. \ref{fig:coopModes} illustrates that improper $p$ and $\tau$
can lead to trust imbalance and reduce the cooperation area, making
trust difficult to be established. Case IV yields the maximal cooperation
area where the dominance between the SP and the client is balanced
in this case.

\vspace{-0.3cm}

\section{Performance Trade-off\label{sec:TradeoffAnalysis}}

\subsection{Transmission Efficiency}

 In this section, we would like to discuss the inherent trade-off
relationship between the transmission efficiency, security integrity,
and cooperation margin in our framework. While the above analysis
is based on a given slot time $t$, the slot time, in fact, has a
significant impact on the trust-building process. Recall that, during
the service process, the client needs to send a packet including the
smart contract, i.e., RMSC, to make the payment in every round. Only
when the SP receives it within the trial time will the remainder service
be delivered. The slot time $t$ affects how frequently the smart
contracts are transmitted. \textcolor{black}{We denote $t_{sc}$ as
the duration to transmit a smart contract, i.e., the communication
over}head. To quantify the impact of the communication overhead, we
define the transmission efficiency $\eta$ as the proportion of time
to transmit payload data, which is given by
\begin{equation}
\eta=1-\frac{t_{sc}}{t}.\label{eq:transEfficiency}
\end{equation}
Apparently, a shorter slot time $t$ results in more frequent interactions
and also a longer communication overhead. Conversely, if we set a
larger slot time $t$, we can reduce the proportion of time to transmit
the smart contract in each round and thus improve the transmission
efficiency, which is suitable for devices with limited bandwidth resources.

\vspace{-0.3cm}

\subsection{Service Integrity\label{subsec:ChannelReliability}}

The proposed trusted access services via repeated interactions may
be interrupted by channel fading, even though both sides are willing
to cooperate. If the SP does not receive the packets containing the
smart contract payment in one round, then the service process is terminated.
A service with the duration $T$ requires $M=\lceil T/t\rceil$ rounds
to finish. If a shorter slot time $t$ is used, then it will take
more rounds to deliver the same service, and the cooperation relationship
is more likely to be terminated. 

Hence, we define the service integrity, denoted by $\zeta$, to quantify
the probability that, under the cooperation conditions, the complete
service is delivered without being terminated. Since the cooperation
conditions are satisfied, the unexpected termination of the service
process can only be caused by channel outage. Consequently, the service
integrity equals the probability that the service requires $M$ rounds
to complete and the packets containing payments are successfully transmitted
in these $M$ rounds, i.e.,
\begin{equation}
\zeta=\sum_{m=1}^{\infty}\left(1-d\right)^{m}\text{Pr}\left(M=m\right).\label{eq:servIntegrity}
\end{equation}
 Specifically, if the service is assumed to obey an exponential distribution
with mean $t_{ave}$,\footnote{$\text{Pr}\left(M=m\right)$ can be derived accordingly if the service
time follows a different distribution.} then we have 
\begin{align}
\text{Pr}\left(M=m\right) & =\text{Pr}\left(\left(m-1\right)t<T\leq mt\right)\nonumber \\
 & =\exp\left(-\frac{\left(m-1\right)t}{t_{ave}}\right)-\exp\left(-\frac{mt}{t_{ave}}\right).\label{eq:P(M=00003Dm)}
\end{align}
By substituting \eqref{eq:P(M=00003Dm)} into \eqref{eq:servIntegrity},
we have
\begin{align}
\zeta= & 1-\frac{d}{1-\left(1-d\right)\exp\left(-t/t_{\text{ave}}\right)}.\label{eq:pnoloss}
\end{align}
 \eqref{eq:pnoloss} indicates that the service integrity $\zeta$
is monotonically decreasing with respect to $d$, implying that the
established cooperation is directly influenced by the channel outage.
Furthermore, a shorter slot time $t$ degrades the service integrity
$\zeta$. It is because a shorter slot time increases the rounds to
deliver a service and raises the risk of payment losses and unexpected
termination, even if both sides are willing to cooperate. Therefore,
a longer slot time $t$ may enhance the cooperation against the channel
unreliability.\vspace{-0.5cm}

\subsection{Cooperation Margin}

For a given service, the slot time $t$ also affects the payoff
of different strategies and thus influences the robustness of the
cooperation. According to Section \ref{sec:coopCond}, ALLD is dominant
among all the strategies involving defection. Hence, we define the
normalized gap between the payoffs of COOP and ALLD as the cooperation
margin of SP and client, which are, respectively, expressed as:
\begin{align*}
\triangle\pi_{s} & =1-\pi_{s}\left(\text{COOP},\text{ALLD}\right)/\pi_{s}\left(\text{COOP},\text{COOP}\right),\\
\triangle\pi_{c} & =1-\pi_{c}\left(\text{COOP},\text{ALLD}\right)/\pi_{c}\left(\text{COOP},\text{COOP}\right).
\end{align*}
The larger the cooperation margin, the more the players can gain from
COOP than defection, making them more inclined to cooperate with each
other. Hence, the cooperation margin can be viewed as a quantitative
measure of the system robustness against misbehavior. Note that $\triangle\pi_{s}\neq\text{\ensuremath{\triangle\pi_{c}}}$
in most cases. However, if $p$ and $\tau$ are optimized according
to \eqref{eq:optw_ptau1}, the normalized payoff gap is the same for
both the SP and the client, and thus, we can drop the subscript and
denote the cooperation margin $\triangle\pi=\triangle\pi_{s}=\triangle\pi_{c}$,
given by
\begin{equation}
\triangle\pi=\frac{\left(1-d\right)w-c/\Gamma\left(d\right)}{1-c/\Gamma\left(d\right)}.\label{eq:robustness-1}
\end{equation}
Under the exponential service assumption, we have $w=\frac{\text{Pr}\left\{ M\geq k+1\right\} }{\text{Pr}\left\{ M\geq k\right\} }=\exp\left(-t/t_{ave}\right)$.
As a result, the cooperation margin is given by
\begin{equation}
\triangle\pi=\frac{\left(1-d\right)\exp\left(-t/t_{ave}\right)-c/\Gamma\left(d\right)}{1-c/\Gamma\left(d\right)}.\label{eq:robustness}
\end{equation}
Essentially, the cooperation margin is the payoff gap between cooperation
and defection. If $\triangle\pi<0$, both sides would choose to defect
and thus the system is vulnerable. $d^{\star}$ and $w^{\star}$ derived
in Section V can be equivalently obtained by setting $\triangle\pi=0$,
which represent the boundary of cooperation. A positive $\triangle\pi$
indicates that the payoff of COOP is higher than the one involving
any defections. A larger cooperation margin implies a more robust
system. Even if some players may be irrational, i.e., they concern
more than payoff, they are still less unlikely to deviate from COOP.
Specifically, \eqref{eq:robustness} shows that $\triangle\pi$ is
monotonically decreasing with respect to $d$, implying that both
sides are more likely to be dishonest in harsher environments. Moreover,
we further find that $\triangle\pi$ is decreasing with respect to
slot time $t$. A shorter slot time $t$ means more service rounds
and thus incentivizes both parties to establish a long-term relationship
by adopting the COOP strategy. 
\begin{figure*}
\begin{raggedright} \centering \subfigure[]{\includegraphics[width=0.33\textwidth]{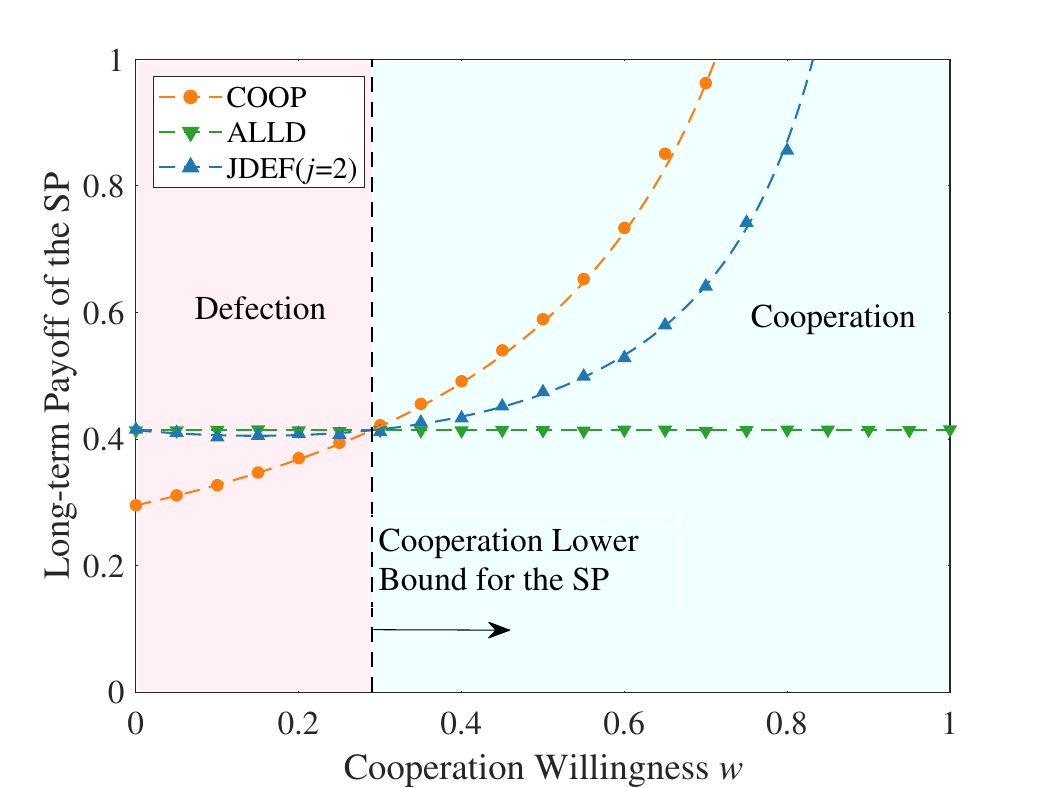}}
\hfill{}\subfigure[]{\includegraphics[width=0.33\textwidth]{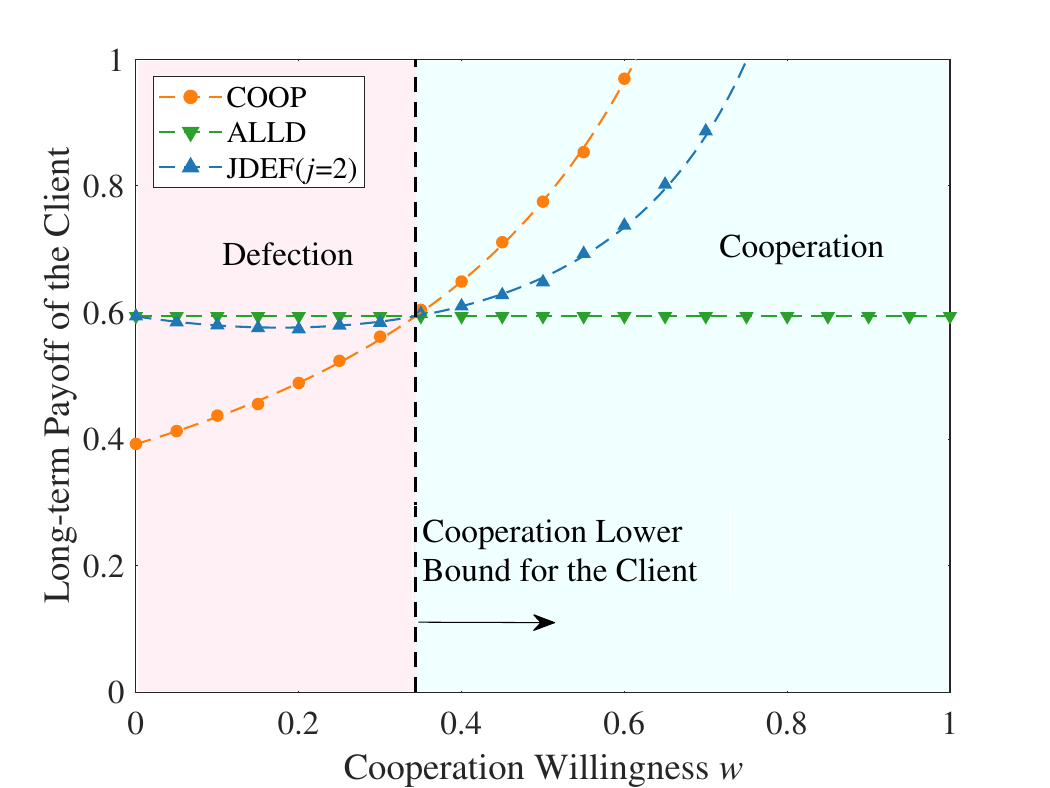}}\hfill{}\subfigure[]{\includegraphics[width=0.33\textwidth]{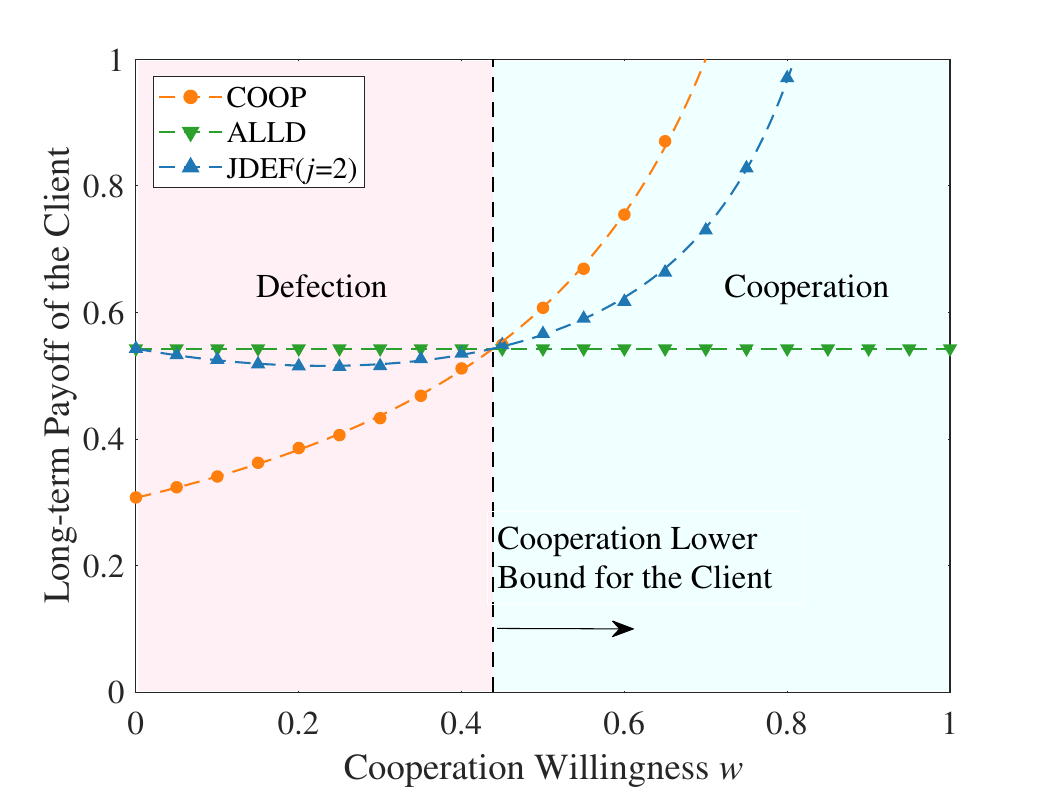}}\end{raggedright}
\centering{}

\caption{Long-term payoffs of the players\textquoteright{} different strategies
for different cooperation willingness $w$. (a) SP. (b) Client with
$\Gamma\left(d\right)=\Gamma\left(0\right)\left(1-d\right)$. (c)
Client with $\Gamma\left(d\right)=\Gamma\left(0\right)\exp\left(-\varphi d\right)$
and $\varphi=10$.\label{fig:simPayoffPerw}}
\vspace{-0.6cm}
\end{figure*}
\begin{figure*}
\begin{raggedright} \centering \subfigure[]{\includegraphics[width=0.33\textwidth]{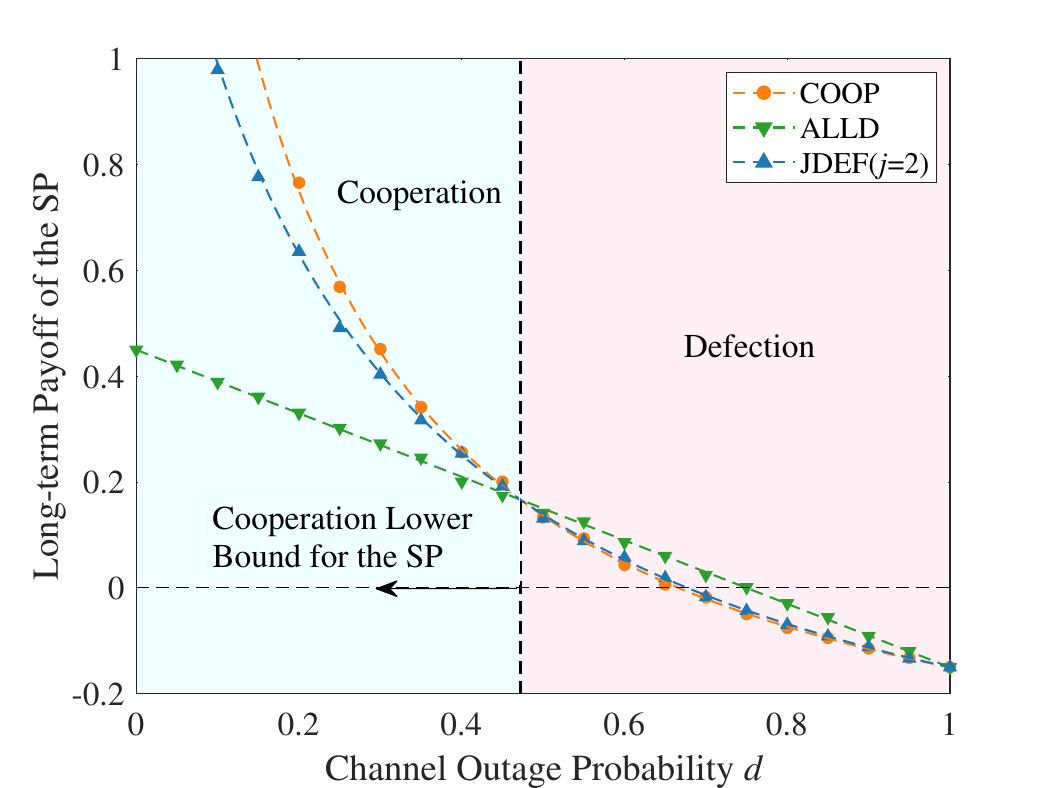}}
\hfill{}\subfigure[]{\includegraphics[width=0.33\textwidth]{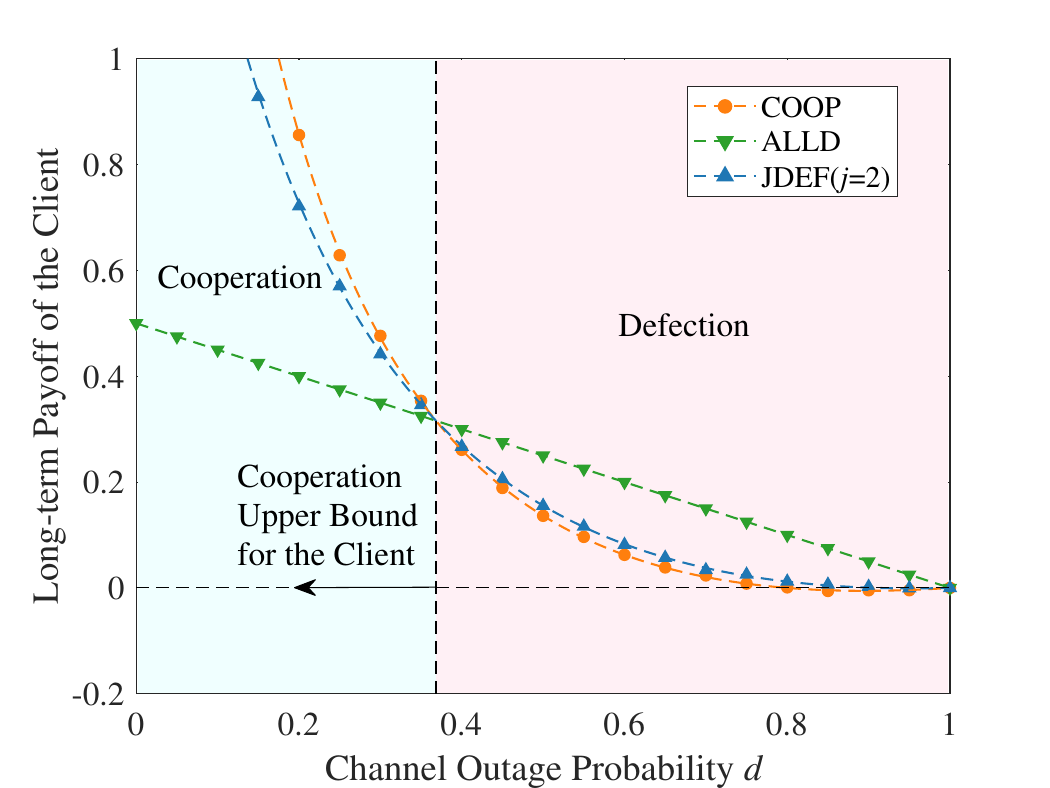}}\hfill{}\subfigure[]{\includegraphics[width=0.33\textwidth]{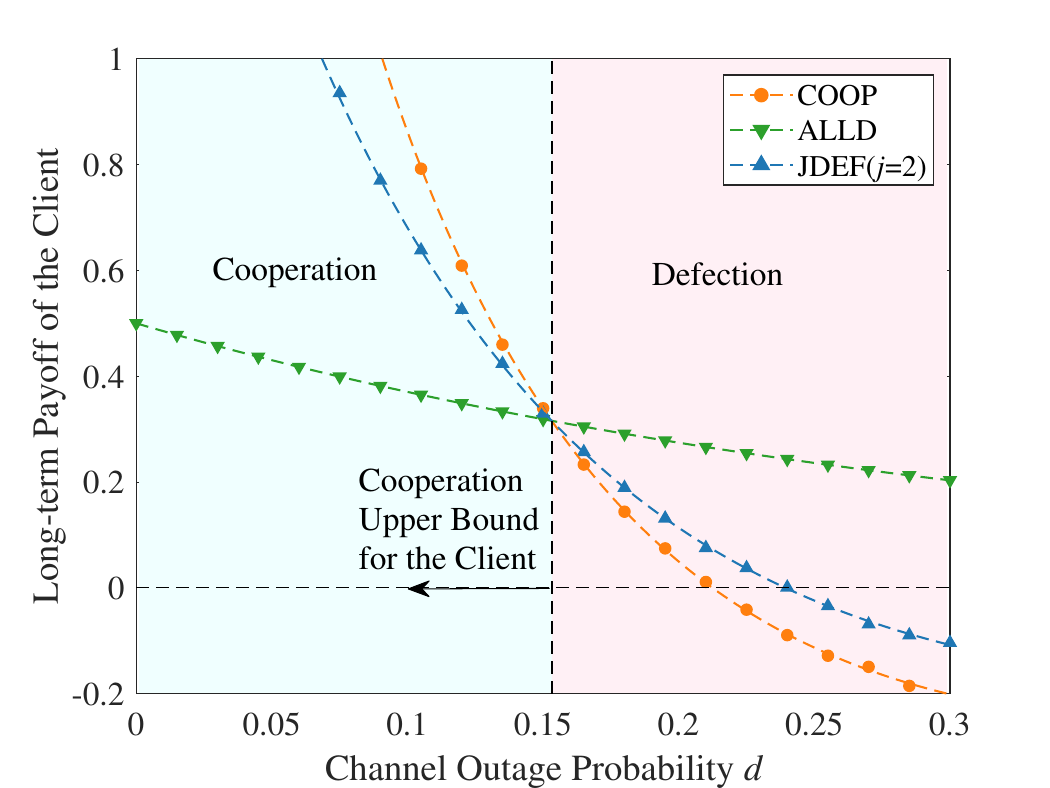}}\end{raggedright}
\centering{}

\caption{Long-term payoffs of the players\textquoteright{} different strategies
for different channel outage probability $d$. (a) SP. (b) Client
with $\Gamma\left(d\right)=\Gamma\left(0\right)\left(1-d\right)$.
(c) Client with $\Gamma\left(d\right)=\Gamma\left(0\right)\exp\left(-\varphi d\right)$
and $\varphi=10$.\label{fig:simpayoffPerd}}
\vspace{-0.6cm}
\end{figure*}

\vspace{-0.3cm}

\subsection{Trade-off Relationship}

Now we analyze our trust-building framework from three different aspects:
transmission efficiency, service integrity, and cooperation margin.
We find that all of them are connected to the slot time $t$. Given
channel outage $d$, average service time $t_{ave}$, and cost $c$,
a shorter slot time $t$ increases the communication overhead and
reduces the transmission efficiency, and meanwhile, it may hurt the
service integrity. Furthermore, a shorter slot time $t$ can enlarge
the cooperation margin and make the participants more inclined to
cooperate. Consequently, one can see the subtle trade-off relationship
among them, which is characterized by the following theorem.
\begin{thm}
\label{thm:tradeoffs}The trade-off relationship between transmission
efficiency $\eta$, cooperation margin $\triangle\pi$ and service
integrity $\zeta$ is given by:
\begin{align}
\triangle\pi & =1-\frac{d}{\left(1-\zeta\right)\left(1-c/\Gamma\left(d\right)\right)},\label{eq:tradeoff1}\\
\triangle\pi & =1-\frac{1-\left(1-d\right)\exp\left(-\frac{t_{sc}}{t_{\text{ave}}}\frac{1}{1-\eta}\right)}{1-c/\Gamma\left(d\right)},\label{eq:tradeoff2}\\
\zeta & =1-\frac{d}{1-\left(1-d\right)\exp\left(-\frac{t_{sc}}{t_{\text{ave}}}\frac{1}{1-\eta}\right)}.\label{eq:tradeoff3}
\end{align}
\end{thm}
\begin{proof}
From \eqref{eq:pnoloss} we have $\exp\left(-\frac{t}{t_{ave}}\right)=\frac{1-\zeta-d}{\left(1-d\right)\left(1-\zeta\right)}$.
Substituting it into \eqref{eq:robustness} yields \eqref{eq:tradeoff1}.
According to \eqref{eq:transEfficiency}, we have $t=t_{sc}/\left(1-\eta\right)$.
\eqref{eq:tradeoff2} and \eqref{eq:tradeoff3} can be obtained by
substituting it into \eqref{eq:robustness} and \eqref{eq:pnoloss},
respectively.
\end{proof}
\begin{figure}
\begin{raggedright} \centering \subfigure[]{\includegraphics[width=0.5\textwidth]{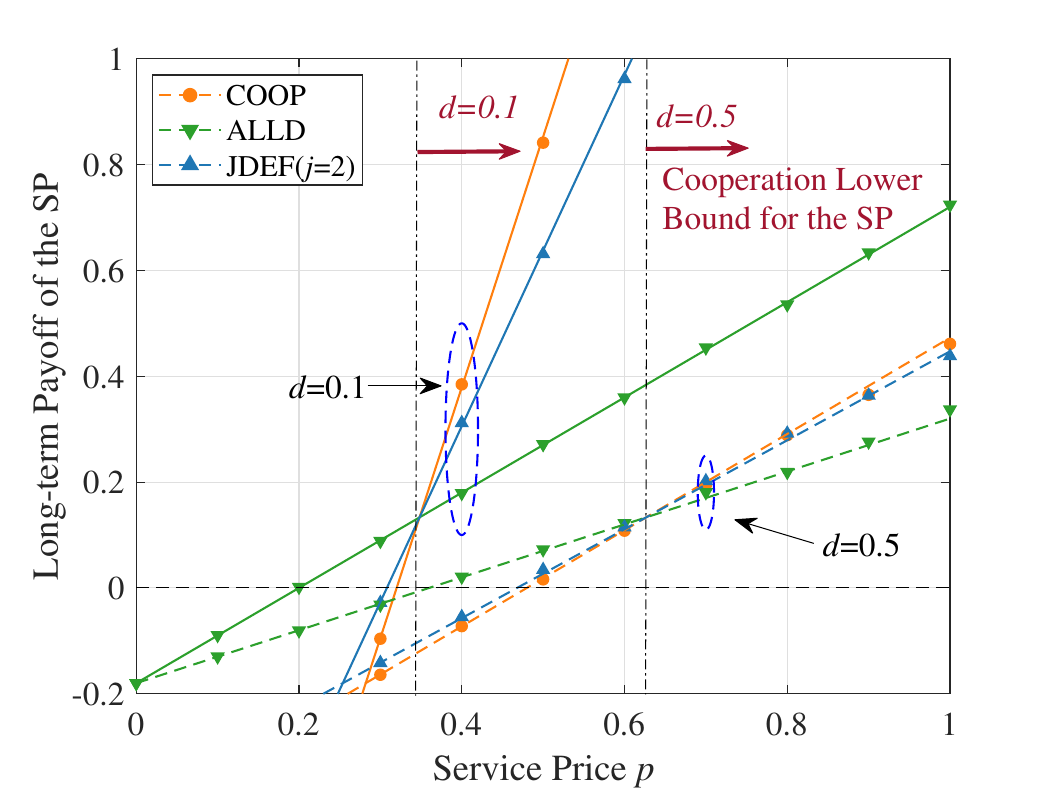}}
\hfill{}\subfigure[]{\includegraphics[width=0.5\textwidth]{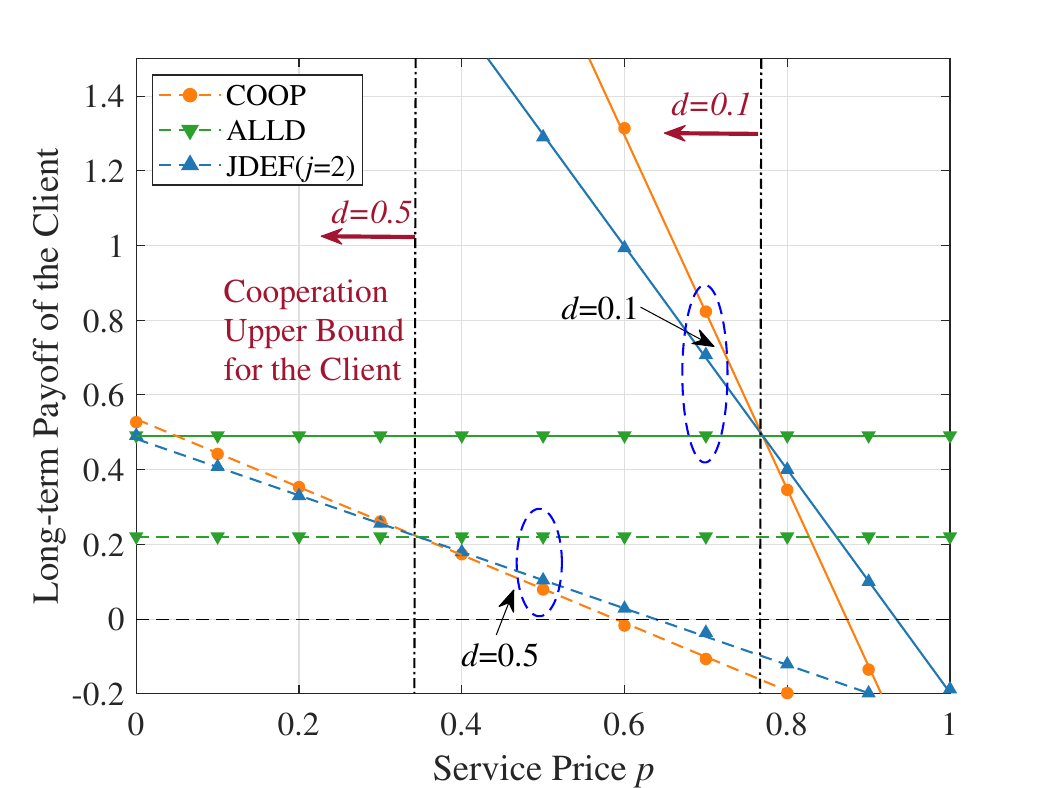}}\end{raggedright}
\centering{}

\caption{Long-term payoffs for different price $p$. (a) SP. (b) Client with
$\Gamma\left(d\right)=\Gamma\left(0\right)\exp\left(-\varphi d\right)$
and $\varphi=2$.\label{fig:simpayoffPerp}}
\vspace{-0.6cm}
\end{figure}
\begin{figure}
\begin{raggedright} \centering \subfigure[]{\includegraphics[width=0.5\textwidth]{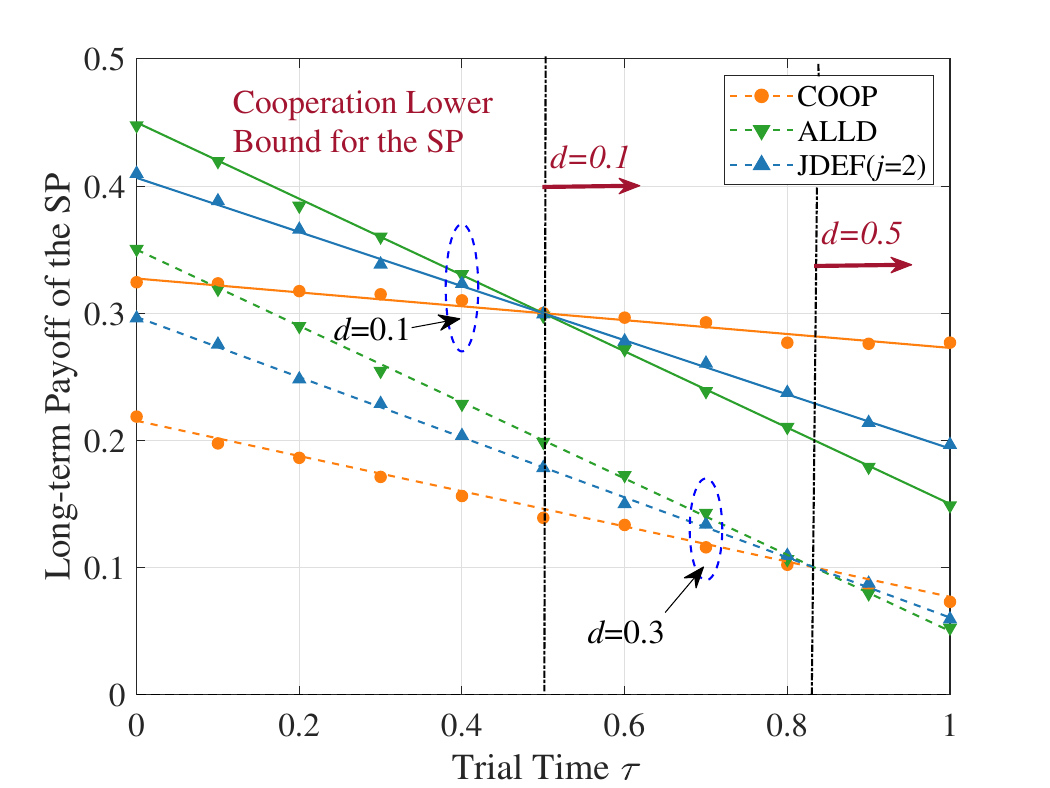}}\hfill{}\subfigure[]{\includegraphics[width=0.5\textwidth]{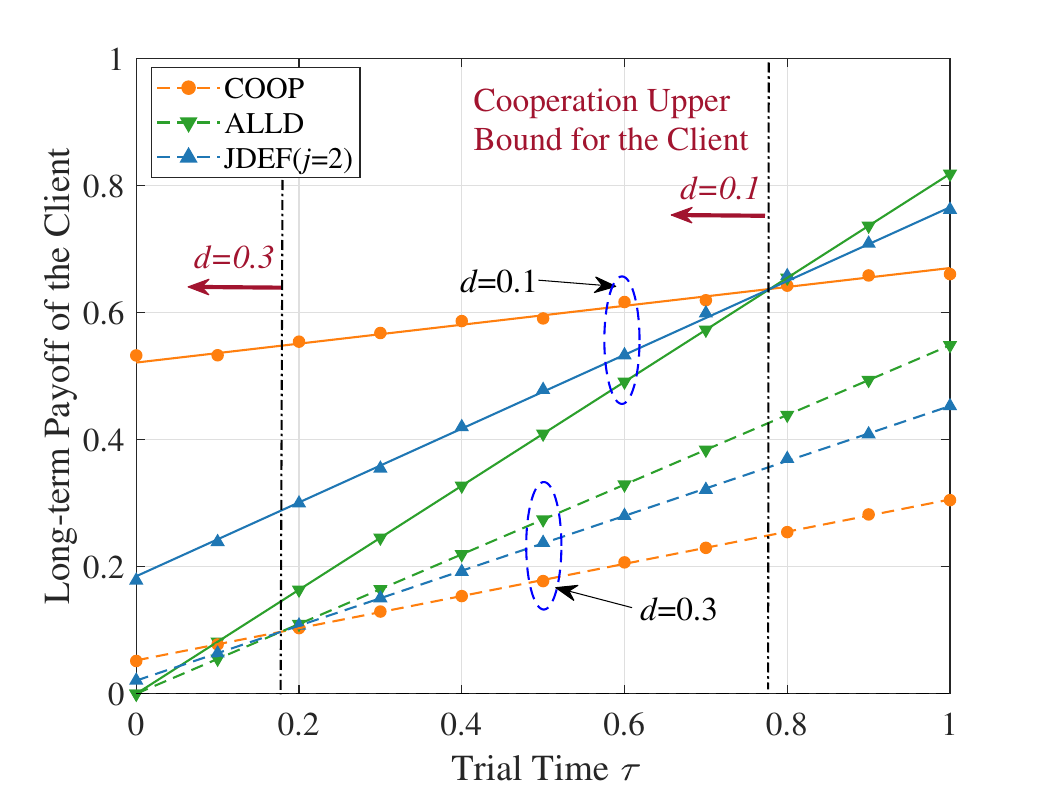}}\end{raggedright}
\centering{}

\caption{Long-term payoffs for different trial time $\tau$. (a) SP. (b) Client
with $\Gamma\left(d\right)=\Gamma\left(0\right)\exp\left(-\varphi d\right)$
and $\varphi=2$.\label{fig:simpayoffPertau}}
\vspace{-0.6cm}
\end{figure}
Theorem \ref{thm:tradeoffs} explicitly reveals the trade-off relationship
of the trust-building process in several different aspects. First,
\eqref{eq:tradeoff1} represents the trade-off between cooperation
margin and service integrity. The better the service integrity $\zeta$
is, the smaller the cooperation margin is. Their trade-off relationship
can be balanced by adjusting the slot time $t$. Second, \eqref{eq:tradeoff2}
shows that, given $d$ and $c$, a higher transmission efficiency
$\eta$, which implies a longer slot time $t$, may decrease the cooperation
margin $\Delta\pi$ and degrade the system robustness. At last, we
can improve the transmission efficiency and enhance the service integrity
by setting a longer slot time $t$. These three metrics can be properly
traded off by using a suitable slot time $t$. Furthermore, we underline
that the slot time $t$ must satisfy $t_{sc}<t<t_{ave}\ln\left(\left(1-d\right)\Gamma\left(d\right)/c\right)$
to guarantee that the smart contract can be transmitted in one round
and also to ensure a positive $\triangle\pi$.

\vspace{-0.3cm}

\section{Simulations\label{sec:Simulations}}

In this section, we present simulation results to support our above
analysis and demonstrate the impact of objective channel uncertainty
and subjective cooperation willingness on the trust-building process.
In this section, analytical and simulation results are represented
by lines and markers, respectively.

Fig. \ref{fig:simPayoffPerw} shows the long-term payoffs for the
SP and the client under different cooperation willingness $w$ with
$d=0.01$. The system parameters are set to $c=0.3$, $p=0.6$, $\Gamma\left(0\right)=1$,
$t=1$, and $\tau=0.6$. We assume three strategies, COOP, ALLD, and
JDEF ($j=2$) with the opponent playing COOP. It can be observed that
ALLD has a constant long-term payoff, while COOP increases monotonically
in $w$. In Fig. \ref{fig:simPayoffPerw}(a), when $w$ meets the
cooperation condition \eqref{eq:nashCondSPw}, the payoff of COOP
is the highest, implying that the SP is more inclined to cooperate
and ensure the required QoS. Similarly, in Fig. \ref{fig:simPayoffPerw}(b)
and \ref{fig:simPayoffPerw}(c), when $w$ satisfies \eqref{eq:nashCondCw},
COOP is more profitable for the client. Fig. \ref{fig:simPayoffPerw}(c)
uses the diminishing marginal utility function, and the cooperation
threshold is higher compared to the linear one in Fig. \ref{fig:simPayoffPerw}(b).

Fig. \ref{fig:simpayoffPerd} illustrates the long-term payoffs of
different strategies against COOP under different outage probability
$d$ with $w=0.9$. The system parameters are set to $c=0.3$, $p=0.6$,
$\Gamma\left(0\right)=1$, $t=1$, and $\tau=0.5$. The results show
that, as $d$ increases, the long-term payoffs for all strategies
reduce. When $d$ is below the cooperation threshold \eqref{eq:nashThreSPd}
and \eqref{eq:nashThreCd}, the payoff of COOP is higher than other
defection strategies, encouraging both the SP and the client to cooperate.
In Fig. \ref{fig:simpayoffPerd}(c), due to the diminishing marginal
utility, the client's payoffs decrease more rapidly with increasing
$d$ compared to the linear utility in Fig. \ref{fig:simpayoffPerd}(b).
The cooperation threshold for $d$ is also lower than in the linear
scenario, implying the cooperation is more difficult to establish
with a diminishing marginal utility.

Fig. \ref{fig:simpayoffPerp} explores the impact of the service price
$p$ on the long-term payoffs with $c=0.3$, $\Gamma\left(0\right)=1$,
$\tau=0.6$, $t=1$ and $w=0.9$. We use the utility function $\Gamma\left(d\right)=\Gamma\left(0\right)\exp\left(-\varphi d\right)$
with $\varphi=2$. In Fig. \ref{fig:simpayoffPerp}(a), for the SP,
as $p$ increases, the long-term payoffs of all the strategies increase.
When $p$ satisfies the cooperation condition \eqref{eq:nashThreSPd},
COOP becomes the most profitable strategy. Fig. \ref{fig:simpayoffPerp}(b)
shows that, for the client, the payoffs decline with increasing $p$,
except for ALLD. When $p$ satisfies \eqref{eq:nashThreCd}, the client
would prefer COOP to obtain more benefits. One can see that the price
needs to be higher for the SP to cooperate but lower for the client.
To enable cooperation, a proper price $p$ has to be set between the
SP\textquoteright s lower bound and the client\textquoteright s upper
bound. Additionally, as $d$ increases, the feasible range of $p$
narrows. When $d$ reaches $0.5$, the lower bound for the SP even
exceeds the upper bound for the client, implying that cooperation
cannot be achieved. 

Fig. \ref{fig:simpayoffPertau} illustrates the impact of trial time
$\tau$ on long-term payoffs, with $c=0.3$, $p=0.5$, $\Gamma\left(0\right)=1$,
$t=1$, and $w=0.5$. The utility function is set to $\Gamma\left(d\right)=\Gamma\left(0\right)\exp\left(-\varphi d\right)$
with $\varphi=2$. In Fig. \ref{fig:simpayoffPertau}(a), as $\tau$
increases, the long-term payoffs for the SP decrease. When the cooperation
condition \eqref{eq:nashThreSPd} is met, the SP is more inclined
to cooperate. Conversely, Fig. \ref{fig:simpayoffPertau}(b) illustrates
that, as $\tau$ increases, the client benefits more but is more inclined
to defect, as the payoff of ALLD grows the fastest. To achieve cooperation,
$\tau$ must be set between the SP\textquoteright s lower bound and
the client\textquoteright s upper bound, and the feasible range of
$\tau$ also narrows as $d$ increases. Cooperation is impossible
when $d$ is 0.5, as the SP\textquoteright s lower bound exceeds the
client\textquoteright s upper bound. 

\begin{figure}
\begin{raggedright} \centering \subfigure[]{\includegraphics[width=0.43\textwidth]{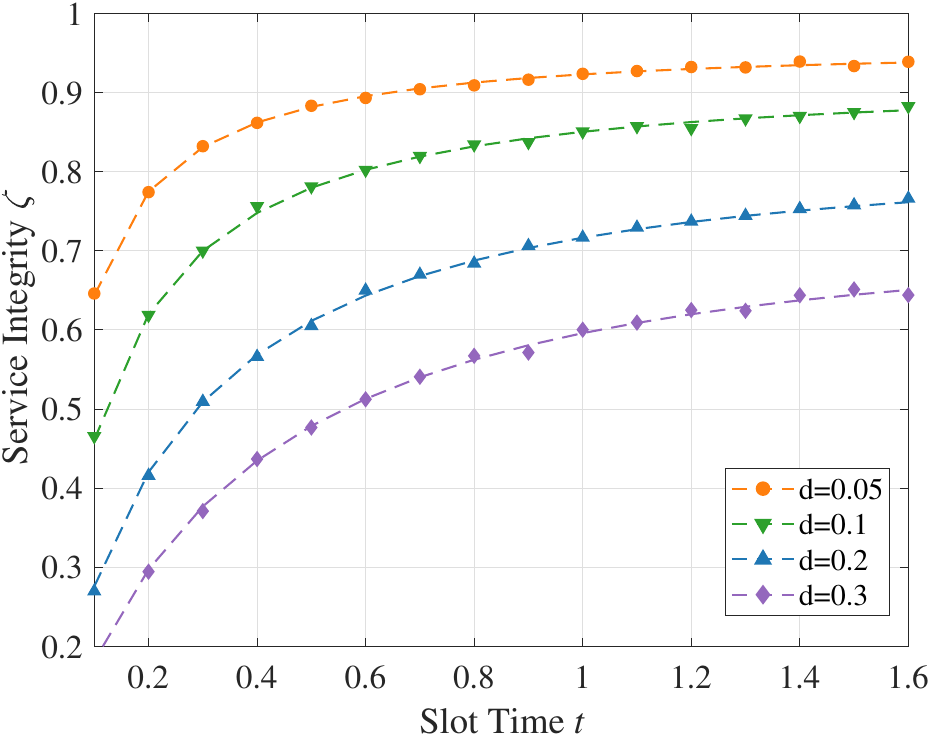}}
\hfill{}\subfigure[]{\includegraphics[width=0.43\textwidth]{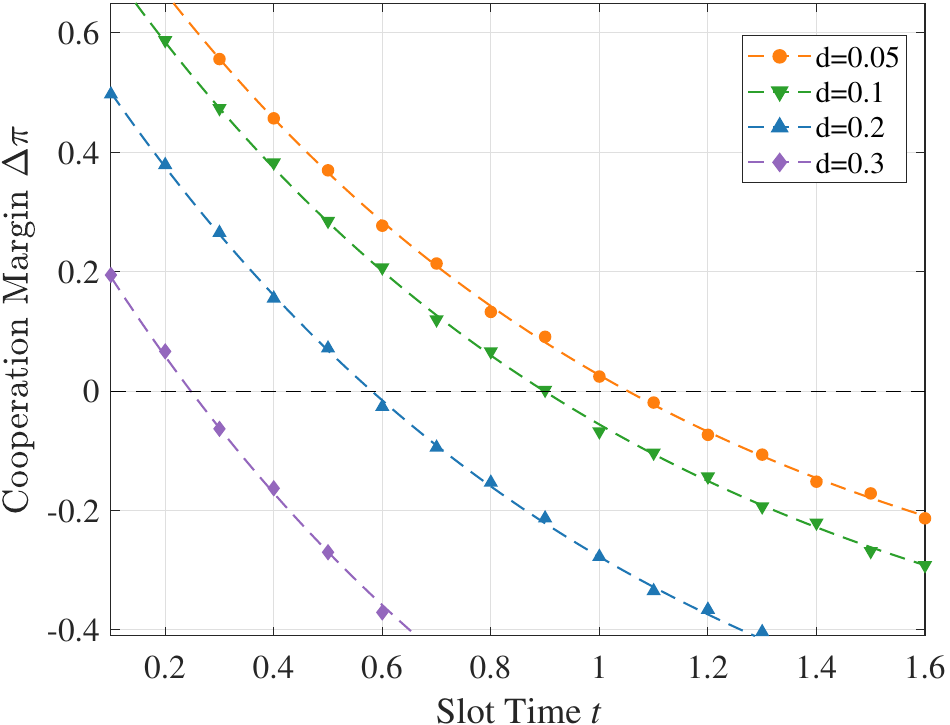}}\end{raggedright}
\centering{}

\caption{The service integrity and cooperation margin for different slot time
$t$. (a) Service integrity. (b) Cooperation margin with $\Gamma\left(d\right)=\Gamma\left(0\right)\exp\left(-\varphi d\right)$
and $\varphi=2$.\label{fig:tradeoff}}
\vspace{-0.6cm}
\end{figure}

Fig. \ref{fig:tradeoff} demonstrates the impact of slot time $t$
on the service integrity and the cooperation margin under different
outage probability $d$. The service time is set to follow the exponential
distribution with $t_{ave}=1$. The trial time $\tau$ and the price
$p$ are optimized according to \eqref{eq:optw_ptau1}. Fig. \ref{fig:tradeoff}
shows the trade-off between the service integrity $\zeta$ and cooperation
margin $\triangle\pi$. As one can see, a high outage probability
$d$ decreases both cooperation margin and service integrity. It also
reveals that, given $d$, a shorter slot time $t$ lowers the transmission
efficiency and the service integrity since it increases the number
of interactions. However, it expands the cooperation margin and makes
both parties more inclined to cooperate. 

\vspace{-0.3cm}

\section{Conclusions\label{sec:Conclusion}}

In this study, we investigated the trusted wireless access framework
based on multi-round interactions and characterized the cooperation
conditions under an unreliable channel. To illustrate the impact of
channel uncertainty on such a framework, we established a repeated
game model by considering channel fading and analyzed the long-term
payoffs of different strategies. We looked into the cooperation conditions
and found that it is more difficult to maintain trusted access services
over an unreliable channel. We pointed out the minimum requirement
on cooperation willingness, under worst channel conditions, and also
investigated the minimum requirement on channel quality, given the
least cooperation willingness of both parties. We found that the above
two design problems can be optimized at the same time. Furthermore,
we introduced the concept of cooperation region to characterize the
framework robustness and considered the cooperation area maximization
for more robust cooperation conditions. We uncovered that both parties
should be taken into consideration in the process of trust-building;
otherwise, the framework would be vulnerable if the establishment
of trust highly relies on one side. We further analyzed transmission
efficiency, security integrity, and cooperative margin, and uncovered
the trade-offs among them. Our results highlighted how channel unreliability
and cooperation willingness affect the trust-building process in the
trusted wireless access framework relying on repeated interactions.
The above results may offer some interesting insights, as a practical
example, to the concept of trust and trustworthiness in the context
of technologies.\vspace{-0.3cm}

\appendix{}

\subsection{Proof of Proposition \ref{thm:7cases}\label{subsec:ProofProposition1}}

We first analyze $d_{\text{min}}\left(w;\tau,p\right)$ according
to the intersection points between $d_{s}\left(w;\tau,p\right)$ and
$d_{c}\left(w;\tau,p\right)$. When $p<2\sqrt{cg_{0}}\sqrt{\frac{\tau}{t}\left(1-\frac{\tau}{t}\right)}$,
they do not intersect. Conversely, when $p>2\sqrt{cg_{0}}\sqrt{\frac{\tau}{t}\left(1-\frac{\tau}{t}\right)}$,
$d_{s}\left(w;\tau,p\right)=d_{c}\left(w;\tau,p\right)$ has two roots,
denoted as $w_{1}$ and $w_{2}$:\vspace{-0.2cm}
\begin{align}
w_{1} & =\left(\frac{p-\sqrt{p^{2}-4c\Gamma\left(0\right)\frac{\tau}{t}\left(1-\frac{\tau}{t}\right)}}{2\frac{\tau}{t}\sqrt{c\Gamma\left(0\right)}}\right)^{2},\label{eq:w1}\\
w_{2} & =\left(\frac{p+\sqrt{p^{2}-4c\Gamma\left(0\right)\frac{\tau}{t}\left(1-\frac{\tau}{t}\right)}}{2\frac{\tau}{t}\sqrt{c\Gamma\left(0\right)}}\right)^{2},\label{eq:w2}
\end{align}
with $d_{\text{min}}\left(w_{1};\tau,p\right)$ and $d_{\text{min}}\left(w_{2};\tau,p\right)$
given by\vspace{-0.1cm}
\begin{align*}
d_{\text{min}}\left(w_{1};\tau,p\right) & =1-\frac{2c\frac{\tau}{t}}{p-\sqrt{p^{2}-4c\Gamma\left(0\right)\frac{\tau}{t}\left(1-\frac{\tau}{t}\right)}},\\
d_{\text{min}}\left(w_{2};\tau,p\right) & =1-\frac{2c\frac{\tau}{t}}{p+\sqrt{p^{2}-4c\Gamma\left(0\right)\frac{\tau}{t}\left(1-\frac{\tau}{t}\right)}}.
\end{align*}
Next, we examine whether the intersection points obtained above are
within the feasible range $0<d,w<1$. When $p<2\sqrt{c\Gamma\left(0\right)}\frac{\tau}{t}$,
$w_{1}<1$ always holds. When $p>2\sqrt{c\Gamma\left(0\right)}\frac{\tau}{t}$,
$w_{1}<1$ yields $p>\sqrt{c\Gamma\left(0\right)}$. Therefore, the
$p\ensuremath{-\tau}$ region corresponding to $w_{1}<1$ can be described
as $p>\max\left(2\sqrt{c\Gamma\left(0\right)}\frac{\tau}{t},\sqrt{c\Gamma\left(0\right)}\right)\cup p<2\sqrt{c\Gamma\left(0\right)}\frac{\tau}{t}$.
In this way, we can divide the $p\ensuremath{-\tau}$ plane into four
regions based on $w_{1}$ and $w_{2}$:
\begin{equation}
\begin{cases}
w_{1}<1, & p>\max\left(2\sqrt{c\Gamma\left(0\right)}\frac{\tau}{t},\sqrt{c\Gamma\left(0\right)}\right)\\
 & \cup p<2\sqrt{c\Gamma\left(0\right)}\frac{\tau}{t};\\
w_{1}>1, & 2\sqrt{c\Gamma\left(0\right)}\frac{\tau}{t}<p<\sqrt{c\Gamma\left(0\right)};\\
w_{2}<1, & p<\min\left(2\sqrt{c\Gamma\left(0\right)}\frac{\tau}{t},\sqrt{c\Gamma\left(0\right)}\right);\\
w_{2}>1, & \sqrt{c\Gamma\left(0\right)}<p<2\sqrt{c\Gamma\left(0\right)}\frac{\tau}{t}\\
 & \cup p>2\sqrt{c\Gamma\left(0\right)}\frac{\tau}{t}.
\end{cases}\label{eq:w0div}
\end{equation}
Similarly, according to the relationship between $d_{\text{min}}\left(w_{1};\tau,p\right)$
and $d_{\text{min}}\left(w_{2};\tau,p\right)$, the $p\ensuremath{-\tau}$
plane can be divided into four regions as follows:
\begin{equation}
\begin{cases}
d_{\text{min}}\left(w_{1};\tau,p\right)<0, & p>\max\left(2c\frac{\tau}{t},\Gamma\left(0\right)\left(1-\frac{\tau}{t}\right)+c\frac{\tau}{t}\right)\\
 & \cup p<2c\frac{\tau}{t};\\
d_{\text{min}}\left(w_{1};\tau,p\right)>0, & 2c\frac{\tau}{t}<p<\Gamma\left(0\right)\left(1-\frac{\tau}{t}\right)+c\frac{\tau}{t};\\
d_{\text{min}}\left(w_{2};\tau,p\right)<0, & p<\min\left(2c\frac{\tau}{t},\Gamma\left(0\right)\left(1-\frac{\tau}{t}\right)+c\frac{\tau}{t}\right);\\
d_{\text{min}}\left(w_{2};\tau,p\right)>0, & \Gamma\left(0\right)\left(1-\frac{\tau}{t}\right)+c\frac{\tau}{t}<p<2c\frac{\tau}{t}\\
 & \cup p<2c\frac{\tau}{t}.
\end{cases}\label{eq:d0div}
\end{equation}
By combining \eqref{eq:w0div} and \eqref{eq:d0div}, the $p\ensuremath{-\tau}$
plane can be divided into seven regions listed in Table \ref{tab:coopModes},
corresponding to seven cases of cooperation areas on the $d-w$ plane. 

\vspace{-0.3cm}

\subsection{Proof of Proposition \ref{thm:optCase237}\label{subsec:ProofProposition2}}

The optimal system configuration pair $\left(\tau,p\right)$ for Cases
II, III and VII can be obtained by solving the corresponding optimization
problems. Take Case II as an example, whose cooperation area is given
by $A_{2}\left(\tau,p\right)=\int_{w_{0}}^{1}d_{s}\left(w;\tau,p\right)\text{d}w$.
$w_{0}$ is the cooperation willingness $w$ satisfying $d_{\text{min}}\left(w_{0};\tau,p\right)=0$,
given by\vspace{-0.1cm}
\begin{align}
w_{0} & =\max\left\{ \frac{c\left(t-\tau\right)}{pt-c\tau},\frac{pt-\Gamma\left(0\right)\left(t-\tau\right)}{\Gamma\left(0\right)\tau}\right\} .\label{eq:w0}
\end{align}
 According to the Leibniz integral rule, the derivative of $A_{2}\left(\tau,p\right)$
with respect to $p$ can be calculated by
\begin{align}
\frac{\partial A_{2}\left(\tau,p\right)}{\partial p} & =\int_{w_{0}}^{1}\frac{\partial d_{s}\left(w,\tau,p\right)}{\partial p}\text{d}w-d_{s}\left(w_{0},\tau,p\right)\frac{\text{\ensuremath{\partial}}w_{0}}{\text{\ensuremath{\partial}}p}.\nonumber \\
 & =\int_{w_{m}}^{1}\frac{1}{p}\left(1-d_{s}\left(w;\tau,p\right)\right)\text{d}w>0.\label{eq:proofCase2}
\end{align}
where $d_{s}\left(w_{0};\tau,p\right)=0$ and $\frac{\partial d_{s}\left(w;\tau,p\right)}{\partial p}=\frac{1}{p}\left(1-d_{s}\left(w;\tau,p\right)\right)$.
Hence, for a fixed $\tau$, $A_{2}\left(\tau;p\right)$ is monotonically
increasing in $p$. As a result, its maximum value must occur at the
boundary $p=\sqrt{c\Gamma\left(0\right)}$. We can find that $A_{2}\left(\tau;p\right)$
is monotonically increasing in $\tau$, with $p=\sqrt{c\Gamma\left(0\right)}$,
and the optimal configuration for Cases II is $\left(\frac{1}{2}t\text{,}\sqrt{c\Gamma\left(0\right)}\right)$.
Similarly, we can prove that the optimal solutions to Cases III and
VII are $\left(\frac{\Gamma\left(0\right)}{\Gamma\left(0\right)+c}t,2\frac{\Gamma\left(0\right)c}{\Gamma\left(0\right)+c}\right)$
and $\left(\frac{\Gamma\left(0\right)-\sqrt{c\Gamma\left(0\right)}}{\Gamma\left(0\right)-c}t,\sqrt{c\Gamma\left(0\right)}\right)$,
respectively. \vspace{-0.3cm}

\subsection{Proof of Proposition \ref{thm:optCase56}\label{subsec:ProofProposition3}}

Take Case\textbf{ }V as an example, whose cooperation area is given
by $A_{5}\left(\tau,p\right)=\int_{w_{0}}^{w_{1}}d_{s}\left(w;\tau,p\right)\text{d}w+\int_{w_{1}}^{1}d_{c}\left(w;\tau;p\right)\text{d}w.$
Since $d_{s}\left(w_{0};\tau,p\right)=0$ and $d_{s}\left(w_{1};\tau,p\right)=d_{c}\left(w_{1};\tau,p\right)$,
the derivative of $A\left(\tau;p\right)$ with respect to $p$ can
be calculated by applying the Leibniz integral rule:
\begin{align}
\frac{\partial A_{5}\left(\tau;p\right)}{\partial p}=\int_{w_{m}}^{w_{1}}\frac{\partial d_{s}\left(w;\tau,p\right)}{\partial p}\text{d}w+\int_{w_{1}}^{1}\frac{\partial d_{c}\left(w;\tau,p\right)}{\partial p}\text{d}w\nonumber \\
=\int_{w_{0}}^{w_{1}}\frac{1}{p}\left(1-d_{s}\left(w;\tau,p\right)\right)\text{d}w+\int_{w_{1}}^{1}\frac{1}{p}\left(d_{c}\left(w;\tau,p\right)-1\right)\text{d}w,\label{eq: A5_1}
\end{align}
where the second equality follows from $\frac{\partial d_{c}\left(w;\tau,p\right)}{\partial p}=\frac{1}{p}\left(d_{c}\left(w;\tau,p\right)-1\right)$.
Define $A_{s/c}\left(\tau,p;x,y\right)\triangleq\int_{x}^{y}d_{s/c}\left(w;\tau,p\right)\text{d}w$.
Since the stationary point of $A_{5}\left(\tau,p\right)$ must satisfy
$\frac{\partial A_{5}\left(\tau,p\right)}{\partial p}=0$, we have
\begin{equation}
A_{c}\left(\tau,p;w_{1},1\right)-A_{s}\left(\tau,p;w_{0},w_{1}\right)=1-2w_{1}+w_{0}.\label{eq:A5}
\end{equation}
Similarly, the optimal $\left(\tau,p\right)$ for Case VI must satisfy
\begin{equation}
A_{s}\left(\tau,p;w_{2},1\right)-A_{c}\left(\tau,p;w_{0},w_{2}\right)=1-2w_{2}+w_{0}.\label{eq:A6}
\end{equation}
Hence, the proposition is proven.

\bibliographystyle{IEEEtran}
\bibliography{IEEEabrv,noisefastaccess_ref}

\begin{thebibliography}{10}
\providecommand{\url}[1]{#1}
\csname url@samestyle\endcsname
\providecommand{\newblock}{\relax}
\providecommand{\bibinfo}[2]{#2}
\providecommand{\BIBentrySTDinterwordspacing}{\spaceskip=0pt\relax}
\providecommand{\BIBentryALTinterwordstretchfactor}{4}
\providecommand{\BIBentryALTinterwordspacing}{\spaceskip=\fontdimen2\font plus
\BIBentryALTinterwordstretchfactor\fontdimen3\font minus \fontdimen4\font\relax}
\providecommand{\BIBforeignlanguage}[2]{{%
\expandafter\ifx\csname l@#1\endcsname\relax
\typeout{** WARNING: IEEEtran.bst: No hyphenation pattern has been}%
\typeout{** loaded for the language `#1'. Using the pattern for}%
\typeout{** the default language instead.}%
\else
\language=\csname l@#1\endcsname
\fi
#2}}
\providecommand{\BIBdecl}{\relax}
\BIBdecl

\bibitem{You2021}
X.~You, C.-X. Wang, J.~Huang, X.~Gao, Z.~Zhang, M.~Wang, Y.~Huang, C.~Zhang, Y.~Jiang, J.~Wang \emph{et~al.}, ``Towards {6G} wireless communication networks: Vision, enabling technologies, and new paradigm shifts,'' \emph{Sci. China Inf. Sci.}, vol.~64, pp. 1--74, Nov. 2021.

\bibitem{Dhiman2024}
P.~Dhiman, N.~Saini, Y.~Gulzar, S.~Turaev, A.~Kaur, K.~U. Nisa, and Y.~Hamid, ``A review and comparative analysis of relevant approaches of zero trust network model,'' \emph{Sens.}, vol.~24, no.~4, p. 1328, Feb. 2024.

\bibitem{Nahar2024}
N.~Nahar, K.~Andersson, O.~Schel{\'e}n, and S.~Saguna, ``A survey on zero trust architecture: Applications and challenges of {6G} networks,'' \emph{IEEE Access}, vol.~12, pp. 94\,753--94\,764, Jul. 2024.

\bibitem{Hajj2021}
S.~Hajj, R.~El~Sibai, J.~Bou~Abdo, J.~Demerjian, A.~Makhoul, and C.~Guyeux, ``Anomaly-based intrusion detection systems: The requirements, methods, measurements, and datasets,'' \emph{Trans. Emerg. Telecommun. Technol.}, vol.~32, no.~4, p. e4240, Apr. 2021.

\bibitem{Mahnamfar2024}
A.~Mahnamfar, K.~Bicakci, and Y.~Uzunay, ``{ROSTAM}: A passwordless web single sign-on solution mitigating server breaches and integrating credential manager and federated identity systems,'' \emph{Comput. Secur.}, vol. 139, p. 103739, Apr. 2024.

\bibitem{Gao2010}
H.~Gao, J.~Yan, and Y.~Mu, ``Dynamic trust model for federated identity management,'' in \emph{Proc. 4th Int. Conf. Network Syst. Secur. (NSS'2010)}, Melbourne, AU, Sep. 2010, pp. 55--61.

\bibitem{Kang2023}
H.~Kang, G.~Liu, Q.~Wang, L.~Meng, and J.~Liu, ``Theory and application of zero trust security: A brief survey,'' \emph{Entropy}, vol.~25, no.~12, p. 1595, Nov. 2023.

\bibitem{Buck2021}
C.~Buck, C.~Olenberger, A.~Schweizer, F.~V{\"o}lter, and T.~Eymann, ``Never trust, always verify: A multivocal literature review on current knowledge and research gaps of zero-trust,'' \emph{Comput. Secur.}, vol. 110, p. 102436, Nov. 2021.

\bibitem{Samaniego2018}
M.~Samaniego and R.~Deters, ``Zero-trust hierarchical management in iot,'' in \emph{Proc. IEEE 12th Int. Congr. Internet Things (ICIOT'18)}, SAN FRANCISCO, CA, USA, Jul. 2018, pp. 88--95.

\bibitem{Sedjelmaci2024}
H.~Sedjelmaci and N.~Ansari, ``Zero trust architecture empowered attack detection framework to secure {6G} edge computing,'' \emph{IEEE Network}, vol.~38, no.~1, pp. 196--202, Jan. 2024.

\bibitem{Liu2024}
Y.~Liu, Z.~Su, H.~Peng, Y.~Xiang, W.~Wang, and R.~Li, ``Zero trust-based mobile network security architecture,'' \emph{IEEE Wireless Communications}, vol.~31, no.~2, pp. 82--88, Apr. 2024.

\bibitem{Baseri2018}
Y.~Baseri, A.~Hafid, and S.~Cherkaoui, ``Privacy preserving fine-grained location-based access control for mobile cloud,'' \emph{Comput. Secur.}, vol.~73, pp. 249--265, Mar. 2018.

\bibitem{Dimitrakos2020}
T.~Dimitrakos, T.~Dilshener, A.~Kravtsov, A.~La~Marra, F.~Martinelli, A.~Rizos, A.~Rosetti, and A.~Saracino, ``Trust aware continuous authorization for zero trust in consumer internet of things,'' in \emph{Proc. IEEE 19th Int. Conf. Trust Securi. Privacy Comput. Commun(TrustCom'20)}, Guangzhou, CN, Dec. 2020, pp. 1801--1812.

\bibitem{Hatakeyama2021}
K.~Hatakeyama, D.~Kotani, and Y.~Okabe, ``Zero trust federation: Sharing context under user control towards zero trust in identity federation,'' in \emph{Proc. IEEE 19th Int. Conf. Pervasive Comput. Commun. Workshops(PerCom Workshops'21)}, Kassel, DE, Mar. 2021, pp. 514--519.

\bibitem{Ward2014}
R.~Ward and B.~Beyer, ``Beyondcorp: A new approach to enterprise security,'' \emph{Login}, vol.~39, no.~6, pp. 6--11, Dec. 2014.

\bibitem{Zheng2018}
Z.~Zheng, S.~Xie, H.-N. Dai, X.~Chen, and H.~Wang, ``Blockchain challenges and opportunities: a survey,'' \emph{Int. J. Web Grid Serv.}, vol.~14, no.~4, pp. 352--375, Oct. 2018.

\bibitem{Ling2019}
X.~Ling, J.~Wang, T.~Bouchoucha, B.~C. Levy, and Z.~Ding, ``Blockchain radio access network ({B-RAN}): Towards decentralized secure radio access paradigm,'' \emph{IEEE Access}, vol.~7, pp. 9714--9723, Jan. 2019.

\bibitem{Le2019}
Y.~{Le}, X.~{Ling}, J.~{Wang}, and Z.~{Ding}, ``Prototype design and test of blockchain radio access network,'' in \emph{Proc. IEEE Int. Conf. Commun. Workshops (ICC'19)}, Shanghai, CN, May 2019.

\bibitem{Ling2020a}
X.~{Ling}, J.~{Wang}, Y.~{Le}, Z.~{Ding}, and X.~{Gao}, ``Blockchain radio access network beyond 5{G},'' \emph{IEEE Wireless Commun.}, vol.~27, no.~6, pp. 160--168, Dec. 2020.

\bibitem{Ling2021}
X.~Ling, Y.~Le, J.~Wang, Z.~Ding, and X.~Gao, ``What is blockchain radio access network?''\hskip 1em plus 0.5em minus 0.4em\relax Hoboken, NJ, USA: Wiley Online Library, Oct. 2021, pp. 1--25.

\bibitem{Ling2021b}
X.~{Ling}, Y.~{Le}, J.~{Wang}, Z.~{Ding}, and X.~{Gao}, ``Practical modeling and analysis of blockchain radio access network,'' \emph{IEEE Trans. Commun.}, vol.~69, no.~2, pp. 1021--1037, Feb. 2021.

\bibitem{cao2024}
W.~Cao, X.~Ling, J.~Wang, Z.~Ding, and X.~Gao, ``A framework for {QoS}-guaranteed fast access services in blockchain radio access network,'' \emph{IEEE Trans. Wireless Commun.}, vol.~23, no.~4, pp. 2711--2725, Apr. 2024.

\bibitem{Antoniou2011}
J.~Antoniou, V.~Papadopoulou, V.~Vassiliou, and A.~Pitsillides, ``Network selection and handoff in wireless networks: A game theoretic approach,'' in \emph{Game Theory for Wireless Communications and Networking}.\hskip 1em plus 0.5em minus 0.4em\relax Boca Raton, FL, USA: CRC Press, Jul. 2011.

\bibitem{Kamhoua2012}
C.~A. Kamhoua, N.~Pissinou, K.~Makki, K.~Kwiat, and S.~S. Iyengar, ``Game theoretic analysis of users and providers behavior in network under scarce resources,'' in \emph{Proc. Int Conf. Comput. Networking Commun. (ICNC'12)}, Maui, HI, USA, Jan. 2012, pp. 1149--1155.

\bibitem{Wang2019}
Y.~Wang, L.~Tian, and Z.~Chen, ``Game analysis of access control based on user behavior trust,'' \emph{Inf.}, vol.~10, no.~4, p. 132, Apr. 2019.

\bibitem{Axelrod1988}
R.~Axelrod and D.~Dion, ``The further evolution of cooperation,'' \emph{Sci.}, vol. 242, no. 4884, pp. 1385--1390, Dec. 1988.

\bibitem{Ling2025}
X.~Ling, Y.~Le, J.~Wang, Y.~Huang, and X.~You, ``Trust and trustworthiness in information and communications technologies,'' \emph{IEEE Wireless Commun.}, accepted, Feb. 2025.

\bibitem{Le2021}
Y.~Le, X.~Ling, J.~Wang, R.~Guo, Y.~Huang, C.~Wang \emph{et~al.}, ``Resource sharing and trading of blockchain radio access networks: {A}rchitecture and prototype design,'' \emph{IEEE Internet Things J.}, vol.~10, no.~14, pp. 12\,025--12\,043, Jul. 2023.

\bibitem{Alevizos2022}
L.~Alevizos, V.~T. Ta, and M.~Hashem~Eiza, ``Augmenting zero trust architecture to endpoints using blockchain: A state-of-the-art review,'' \emph{Secur. Privacy}, vol.~5, no.~1, p. e191, Feb. 2022.

\bibitem{Papadis2020}
N.~Papadis and L.~Tassiulas, ``Blockchain-based payment channel networks: Challenges and recent advances,'' \emph{IEEE Access}, vol.~8, pp. 227\,596--227\,609, Dec. 2020.

\bibitem{Besanko2020}
D.~Besanko and R.~Braeutigam, \emph{Microeconomics}.\hskip 1em plus 0.5em minus 0.4em\relax Hoboken, NJ, USA: Wiley, May 2020.

\bibitem{Wu1995}
J.~Wu and R.~Axelrod, ``How to cope with noise in the iterated prisoner's dilemma,'' \emph{J. Conflict Resolut.}, vol.~39, no.~1, pp. 183--189, Mar. 1995.

\bibitem{Fudenberg1991}
D.~Fudenberg and J.~Tirole, \emph{Game theory}.\hskip 1em plus 0.5em minus 0.4em\relax Cambridge, MA, USA: MIT press, 1991.

\end{thebibliography}

\end{document}